\documentclass[]{aastex631}
\received{January 8, 2024}
\revised{August 14, 2024}
\accepted{August 23, 2024}

\begin{document}

\title{Revealing Potential Initial Mass Function variations with
  metallicity: JWST observations of young open clusters in a
  low-metallicity environment}

\correspondingauthor{Chikako Yasui}
\email{ck.yasui@gmail.com}

\author[0000-0003-3579-7454]{Chikako Yasui}
\affiliation{National Astronomical Observatory of Japan, 2-21-1 Osawa, Mitaka, Tokyo 181-8588, Japan}

\author[0000-0003-1604-9127]{Natsuko Izumi}
\affiliation{Institute of Astronomy and Astrophysics, Academia Sinica, No. 1, Section 4, Roosevelt Road, Taipei 10617, Taiwan}

\author[0000-0003-0769-8627]{Masao Saito}
\affiliation{National Astronomical Observatory of Japan, National Institutes of Natural Sciences, 2-21-1 Osawa, Mitaka, Tokyo 181-8588, Japan}
\affiliation{Department of Astronomical Science, School of Physical Science, SOKENDAI (The Graduate University for Advanced Studies), 2-21-1 Osawa, Mitaka, Tokyo 181-8588,
Japan}

\author[0000-0003-0778-0321]{Ryan M. Lau}
\affiliation{NSF's National Optical-Infrared Astronomy Research Laboratory, 950 North Cherry Avenue, Tucson, AZ 85719, USA}

\author[0000-0003-4578-2619]{Naoto Kobayashi}
\affiliation{Institute of Astronomy, School of Science, University of Tokyo, 2–21–1 Osawa, Mitaka, Tokyo 181–0015, Japan}
\affiliation{Kiso Observatory, Institute of Astronomy, School of Science, University of Tokyo, 10762–30 Mitake, Kiso-machi, Kiso-gun, Nagano 397–0101, Japan}

\author[0000-0001-5644-8330]{Michael E. Ressler}
\affiliation{Jet Propulsion Laboratory, California Institute of Technology \\
4800 Oak Grove Drive, Pasadena, CA 91109, USA}



\begin{abstract}

We present the substellar mass function of star-forming clusters
($\simeq$0.1 Myr old) in a low-metallicity environment ($\simeq$$-$0.7
dex).
We performed deep JWST/NIRCam and MIRI imaging of two star-forming
clusters in Digel Cloud 2, a star-forming region in the Outer Galaxy
($R_G \gtrsim 15$ kpc).
The very high sensitivity and spatial resolution of JWST enable us
to resolve cluster members clearly down to a mass detection limit of
0.02 $M_\odot$, 
enabling the first detection of brown dwarfs in low-metallicity
clusters.
Fifty-two and ninety-one sources were extracted in mass-$A_V$-limited
samples in the two clusters, from which Initial mass functions (IMFs)
were derived by model-fitting the F200W band luminosity function,
resulting in IMF peak masses (hereafter $M_C$) {$\log M_C / M_\odot
  \simeq -1.5 \pm 0.5$} for both clusters.
Although the uncertainties are rather large, the obtained $M_C$ values
are lower than those in any previous study ($\log M_C / M_\odot \sim
-0.5$).
Comparison with the local open clusters with similar ages to the
target clusters ($\sim$$10^6$--$10^7$ yr) suggests a metallicity
dependence of $M_C$, with lower $M_C$ at lower metallicities,
while the comparison with globular clusters, similarly low
metallicities but considerably older ($\sim$$10^{10}$ yr), suggests
that the target clusters have not yet experienced significant
dynamical evolution and remain in their initial physical condition.
The lower $M_C$ is also consistent with the theoretical expectation of
the lower Jeans mass  due to the higher gas density under such low
metallicity. 
The $M_C$ values derived from observations in such an environment
would place significant constraints on the understanding of star
formation.

\end{abstract}



\section{Introduction} \label{sec:intro}

The concept of the Initial Mass Function (IMF) was first introduced by
\citet{Salpeter1955},
who found that a power-law IMF ($d N/d \log m \propto m^{-\Gamma}$)
with a ``Salpeter index'' of $\Gamma = 1.35$ fit the stellar mass
distribution in the solar neighborhood, where $m$ is the mass of a
star and $N$ is the number of stars in some logarithmic mass bins.
Subsequent improvements in observational instrumentation had led to the
detection of turnover in IMFs in various regions, leading to
discussion of substellar IMFs (references in \citealt{Andersen2008}).
In particular, the turnover mass, called the characteristic mass ($M_C$),
and its behavior with various star-forming conditions can provide
valuable insights into the physical processes that regulate star and
brown dwarf formation \citep{Luhman2000}.
In the solar neighborhood, it has been observationally determined to be
approximately constant ($\log M_C /M_\odot \sim -0.5 \pm 0.5$, or $M_C
\sim 0.3$ $M_\odot$) in diverse conditions \citep{Elmegreen2008},
but no models have fully explained why it can be the same.
Theoretical studies point out that only massive stars can form in the
first generation of stars due to inefficient cooling in extreme
conditions (e.g., quite low-metallicity environment such as $Z
\lesssim 10^{-5}$; \citealt{Omukai2005}), and more recently, $M_C$
differences for less extreme cases of metallicities have been
discussed (e.g., \citealt{Chon2021}).
However, there is little discussion as to how low at metallicity $M_C$
begins to change.

For low-metallicity environments, globular clusters have received the
most attention so far.
Globular clusters have low-metallicity of $\sim$$-$0.5 to $-$2 dex
\citep{Paresce2000} and they are generally well-studied due to a long
history of optical observations.
$M_C$ for the clusters has been found to be similar to other
regions in the solar neighborhood \citep{Elmegreen2008}. 
On the other hand, it has been pointed out that low-mass stars may not
be kept intact in the clusters due to the dynamical evolution in the
old systems ($\sim$$10^{10}$ yr), and some suggest slightly higher
$M_C$ for globular clusters compared to other local regions \citep{De
  Marchi2010}. 
Also note that, due to the old ages, the mass range for which the IMF
is derived is very narrow ($\simeq$0.1--1 $M_\odot)$.
This is because high-mass and intermediate-mass stars have already
completed their short lives, and low-mass stars such as brown dwarfs
have completed hydrogen fusion and become cold and dark, making them
difficult to detect.

The Large and Small Magellanic Clouds (LMC/SMC) were targeted for IMF
studies of young clusters in low-metallicity environments.
However, due to the large distance of the LMC/SMC ($\sim$50 kpc), even
with the Hubble Space Telescope, the detection limit had only reached
down to $\sim$1 $M_\odot$ and never reached the $M_C$ (e.g.,
\citealt{Sirianni2000}).
For alternative targets, we have explored the outer Galaxy, which have
generally low metallicities similar to the LMC/SMC (${\rm [O/H]}
\lesssim -0.5$ dex), but have significantly smaller distances ($D
\simeq$5--10 kpc).
 Subaru near-infrared (NIR) imaging of the star-forming regions Digel
Cloud 2 \citep{{Yasui2006},{Yasui2008}}, S207 \citep{Yasui2016b}, S208
\citep{Yasui2016a}, S127 \citep{Yasui2021}, and S209 \citep{Yasui2023},
 all selected as the region where low metallicities have actually been
 determined (e.g., \citealt{Rudolph2006}), 
 has detected stars with masses down to about 0.1 $M_\odot$ were
 detected in all regions, just covering the peak of the IMF.
 For these regions, we primarily estimated the age and IMF at the same
 time, and suggested that IMFs are not significantly different from
 the canonical IMF.
Only for S209 was the detailed derivation of the IMF attempted,
suggesting it had a lower $M_C$ ($<$0.1 $M_\odot$) than those
generally seen in the solar neighborhood ($\sim$0.3 $M_\odot$),
and had a slightly flatter high-mass slope ($\Gamma \simeq -1.0$), IMF
slope for lower masses than the first break mass, compared to the
Salpeter IMF ($\Gamma = -1.35$).
However, because the mass detection limit did not reach masses
sufficiently lower than the $M_C$ (it only reached up to the first
break mass), the actual value could not be determined.

For the next step, we obtained NIR and mid-infrared (MIR) images using
JWST NIRCam/MIRI \citep{Izumi2024}, which has the highest
sensitivities in both NIR/MIR among existing telescopes.
Among star-forming regions in the outer Galaxy, Digel Cloud 2 was
selected as the best target for derivation of substellar IMF because
the derivation of IMFs for young objects requires independent
information about the age ($\sim$0.1 Myr),
which has been well studied for this region previously
\citep{Kobayashi2008}.
In addition, the metallicity of this region has been determined to be
sufficiently low ($\simeq$$-$0.7 dex) by various methods, from the
radio molecular lines of molecular clouds \citep{{Lubowich2004},
  {Ruffle2007}} and optical recombination lines of an associated early
type star (\citealt{Smartt1996}; \citealt{Rolleston2000}), which is
another reason for selecting this region.
Even for the star-forming region located in the outermost region of
the Galaxy ($D \gtrsim 10$ kpc; $R_G \gtrsim 15$ kpc), the very high
sensitivity of JWST allows us to reach the substellar mass regime
($\sim$0.01 $M_\odot$).

In this paper, we derive the IMF by using NIRCam imaging data in the
short wavelength channel, which is the most sensitive wavelength
ranges for young stars.
In future papers, we will report more details on star formation in
this region and discuss in particular the evolution of protoplanetary
disks.
This paper is organized as follows.
In Section~2, we summarize previous studies of Cloud 2, focusing on
studies of star-forming activities, and describe NIR deep imaging
observations of the regions and data reduction with JWST/NIRCam.
Section~3 describes the NIR imaging results for the two young clusters
in Cloud 2.
In Section 4, we construct model F200W band luminosity functions
(F200W-LFs), which are used to derive the IMF, and present the
best-fit IMFs obtained for the Cloud 2 clusters.
In Section~5, we discuss the uncertainties in in the derived $M_C$ for
Cloud 2 clusters, compare the derived $M_C$ with previous studies, and
then discuss a possible metallicity dependence of the $M_C$.
Finally, we conclude in Section~6.

\section{Target, Observations and Data Reduction} \label{sec:Obs}

\subsection{Digel Cloud 2 clusters} \label{subsec:target}

Digel Cloud 2 (hereafter, ``Cloud 2'') was originally discovered as
the largest and most distant molecular cloud in the CO molecular cloud
survey of the Galactic anticenter ($l=130$--155$\arcdeg$) by
\citet{Digel1994}.
The cloud is located in the direction of Galactic longitude
$l=137.75^\circ$ (Galactic latitude $b= -1.00^\circ$) with the LSR
(local standard of rest) velocity ($v_{\rm LSR} = -102.4$ km
s$^{-1}$).  \citet{Digel1994} estimated the distance of the cloud as
$R_G = 28$ kpc (heliocentric distance $D = 21$ kpc).  \citet{Stil2001}
presented \ion{H}{1} observations that revealed a large expanding
supernova remnant (SNR) shell
(the center velocity of $v_{\rm LSR} = -94.2 \pm 0.5$ km s$^{-1}$),
GSH 138-01-94, that is associated with the Cloud 2.  From the
observations, they estimated the distance as $R_G = 23.6$ kpc ($D =
16.6$ kpc).
However, the distance to the associated early-type star MR-1
(\citealt{Muzzio1974}; $v_{\rm LSR} = -102.7 \pm 12$ km s$^{-1}$ by
\citealt{Russeil2007}) is estimated to have smaller values.
From high-resolution optical spectra, \citet{Smartt1996}
estimated it as $R_G = 15$--19 kpc ($D = 8$--12 kpc),
while \citet{Russeil2007} re-estimated $R_G = 14.3 \pm 0.5$ kpc ($D =
6.78 \pm 0.59$ kpc).

As for star formation activity, \citet{de Geus1993} found an
\ion{H}{2} region associated with Cloud 2 ($v_{\rm LSR} = -101$ km
s$^{-1}$) that is excited by MR-1, while \citet{KT2000} found
associated red infrared sources using the QUIST (Quick Infrared
  Survey Telescope), a NIR camera and 25.4 cm Cassegrain telescope, 
to confirm the star-forming activity in this cloud.
Using the QUick InfraRed Camera (QUIRC) at the University of Hawaii
2.2 m telescope, \citet{Kobayashi2008} found two young embedded star
clusters located in the northern and southern dense cores of the
cloud, the Cloud 2-N and -S clusters, respectively.
Using multiwavelength data, they presented clear evidence of sequential
star formation triggered by the SNR, GSH 138-01-94.
They suggest that the compression of the molecular cloud where the
Cloud 2 clusters are embedded was brought on by the \ion{H}{1} shell
expansion, which is well supported from JWST observations
\citep{Izumi2024}.
\citet{Kobayashi2008} also estimated the upper limit of the age of the
cluster at 0.4 Myr from the projected angular difference of the
cluster from the current SNR shell front, using the distance of $R_G =
19$ kpc ($D = 12$ kpc; \citealt{Smartt1996}).
From deeper NIR imaging using the Subaru 8.2-m telescope,
\citet{{Yasui2006},{Yasui2008},{Yasui2009}} discussed properties of
the Cloud 2 clusters.

Recently, the Gaia Data Release 3 (Gaia DR3; \citealt{Gaia2023}) has
provided measurements of the distances to distant objects.
The parallax for MR-1 was obtained with relatively high accuracy, with
the parallax-to-error ratio of this source being more than 10, while
those for the YSOs thought to be associated with the Cloud 2 in
\citet{Kobayashi2008} (IRS 1--7) were not obtained with such high
accuracy.
MR-1 was matched with the Gaia source ID  461019899768797824 
($p = 0.0992 \pm 0.0193$ mas). 
\citet{Bailer-Jones2021} estimated geometric and photogeometric
distances by Bayesian approach using priors based on three-dimensional
model of our Galaxy.
Among two types of distances, the geometric distance,
7.9$^{+1.2}_{-1.1}$ kpc, is adopted for this target because its
parallax SNR is relatively high ($>$5) and the source seems to be in a
crowded region judging from DSS (Digitized Sky Survey) images. 
Because the Gaia astrometric distance is the most reliable at this
time, this value is adopted in this paper.
With this adopted distance, the Galactic radius is 14.8 kpc.
We also note that the upper limit of the Cloud 2 cluster age,
calculated in \citet{Kobayashi2008} was under the assumption of the
distance of 12 kpc, and we now estimate the upper limit to be 0.1 Myr
using the adopted distance here, 7.9 kpc, because the expansion radius
is in proportion to $t^{2/7}$ \citep{Stil2001}.
Therefore, we adopt the clusters' ages of $\simeq$0.1 Myr in this
paper, even considering the uncertainties of the astrometric distance.

According to the standard metallicity curve, metallicity at the
Galactic radius of Cloud 2 is estimated to be $\sim$$-$1 dex (e.g.,
\citealt{Smartt1997}).
However, it should be noted that there is a large variation in the
curve, and some regions exist in environments where the metallicity is
actually not as low as predicted, even though they are located in the
outer Galaxy.
The metallicity of Cloud 2 is therefore independently estimated at
$-$0.7 dex by using radio molecular lines \citep{{Lubowich2004},
  {Ruffle2007}};
this is consistent with the metallicity of MR-1 as measured by
optical high-resolution spectroscopy ($-$0.5 to $-$0.8 dex;
\citealt{Smartt1996}; \citealt{Rolleston2000}).
This metallicity is comparable to that of the LMC ($-$0.5 dex at $D
\sim 50$ kpc) and SMC ($-$0.9 dex at $D \sim 60$ kpc), but given that
the distance is approximately 1/5th of those, it shows that the Cloud
2 is a good target for examining individual, very low mass stars at
subsolar metallicity.

\subsection{JWST NIRCam/MIRI imaging} \label{subsec:nircam_imaging}

The JWST NIRCam and MIRI observations of Cloud 2 were acquired on
January 17, 2023 (UT) as part of the Cycle 1 Guaranteed Time
Observation (GTO) Program (Program ID: 1237, PI: Michael Ressler)
Cloud 2-N was observed with NIRCam in six filters from 1 to 5 $\mu$m
(F115W, F150W, F200W, F356W, F444W, and F405N) with total exposure
times of 344 s each, and with MIRI in three filters from 6 to 25
$\mu$m (F770W, F1280W, and F2100W) with total exposure times of 389 s
each.
Cloud 2-S was observed with NIRCam and MIRI in the same filters as
Cloud 2-N, but the total exposure time was 429 s for NIRCam and
$\simeq$400 s for MIRI (400 s for F770W and F1280, and 411 s for
F2100W).
The observations themselves are summarized in \citet{Izumi2024}. 
In this paper, we primarily use the NIRCam images in the shorter
wavelength filters (F115W, F150W, and F200W)
because they have the highest sensitivities that allow us to derive
IMFs covering the lowest mass objects possible with our JWST
observations.

The NIRCam instrument has two modules (module A and B).
For the short wavelength channel (0.6--2.3 $\mu$m), each module has
four 2K detectors with a pixel scale of 0.031$''$ pixel$^{-1}$, while
each module has one 2K detectors with a pixel scale of 0.063$''$
pixel$^{-1}$ in the long wavelength channel (2.4--5.0 $\mu$m).
The detectors in the short wavelength channel are arranged in a $2
\times 2$ array covering an area of $2.2' \times 2.2'$ with 4$''$--5$''$
gaps between detectors.
In these observations, only module B was used due to the limited
observing periods.
The MIRI instrument uses three 1K arrays, one of which is used for
imaging, covering $74'' \times 113''$ with a pixel scale of 0.11$''$
pixel$^{-1}$.
The field of view for MIRI imaging was set to cover almost the same
area as NIRCam imaging for Cloud 2-N, 
while half the area of NIRCam imaging was covered with MIRI imaging for
Cloud 2-S (see \citealt{Izumi2024} for these details).

Only one field of view of module B was required to cover Cloud 2-N,
while one and a half fields in the long north-south direction covered
Cloud 2-S.
The fields of view were set so that the entire region of the Cloud 2
clusters, identified in \citet{{Yasui2006}, {Yasui2008}, {Yasui2009}},
would be placed in one chip (not falling into the gap).
For the Cloud 2-N cluster, the center of detector B4 was set to the
coordinates of {$(\alpha_{\rm 2000}, \delta_{\rm 2000}) = (02^{\rm h}
  48^{\rm m} 42\fs0, +58^\circ 28' 55\farcs3)$} with position angle
(PA) of telescope V3 axis of 86 deg.
The Cloud 2-S cluster was covered by two mosaics: the center of
detector B4 was set to the coordinates of {$(\alpha_{\rm 2000},
  \delta_{\rm 2000}) = (02^{\rm h} 48^{\rm m} 27\fs2, +58^\circ 23'
  09\farcs8)$} with ${\rm PA} = 87$ deg, and that of the detector B2
was set to $(\alpha_{\rm 2000}, \delta_{\rm 2000}) = (02^{\rm h}
48^{\rm m} 28\fs5, +58^\circ 23' 22\farcs2)$ with the same PA.

Because the stellar density of cluster regions is generally high and
we prefer to use point-spread function (PSF) fitting for the
photometry (Section~\ref{subsec:reduction_photometry}),
NIRCam subpixel dithers were adopted to improve the spatial resolution
of the final combined image mosaic of all the exposures. 
As a result, the combined images have ${\rm FWHM} = 0\farcs07$ for the
F200W band, corresponding to spatial resolution of 550 au at the
distance of 7.9 kpc.

\subsection{Data Reduction and Photometry} \label{subsec:reduction_photometry}
All the data in each band are reduced with the standard procedure
using JWST Science Calibration pipeline version number 1.9.5.
Default values were used for the pipeline parameters except for
``abs\_refcat'', which sets the astrometric reference catalog in Step 3.
'GAIADR2' was set to query the GAIA-based Astrometric Catalog web
service to generate a catalog of all astrometrically measured objects
in the combined field of view of the input image set in order to
improve the astrometry.
We also performed a custom background subtraction step since we do not
have dedicated background data and nebulosity fills many of the fields
of view. We estimated the background of each frame by measuring the
level in several small, visually-selected regions that are free of
nebulosity and then subtracting that value from the frame before
combining the mosaic images.
We constructed NIR pseudocolor images of Cloud 2-N and -S by combining
the NIRCam images for the F115W (blue), F150W (green), and F200W (red)
bands (Figures~\ref{fig:3col_CL2N} and ~\ref{fig:3col_CL2S},
respectively).

From the Stage 3 images, we obtained NIRCam photometry for Cloud 2-N
and Cloud 2-S by fitting the PSF using IRAF\footnote{NOIRLab IRAF is
distributed by the Community Science and Data Center at NSF NOIRLab,
which is managed by the Association of Universities for Research in
Astronomy (AURA) under a cooperative agreement with the U.S. National
Science Foundation.}/DAOPHOT.
To derive the PSF, we selected stars that were bright but not
saturated, that were not close to the edge of the frame, and that did
not have any nearby stars with magnitude differences $<$4 mag.
We performed the PSF photometry in two iterations using the ALLSTAR
routine: the first used the original images, and the second used the
images remaining after the sources from the first iteration had been
subtracted.
We obtained PSF-fit radii of 1.8, 1.9, and 2.3 pixels for the full
widths at half maximum in the F115W, F150W, and F200W bands,
respectively, and set the inner radii and the widths of the sky annuli
to be four and three times the PSF-fit radii, respectively.
Because the flux calibration and zero points were in the process of
being established until recently, stars in the field that were also
detected with Subaru/MOIRCS in our previous study \citep{Yasui2009}
were used as photometric standards, after converting the magnitudes in
the Mauna Kea Observatory (MKO) filter system \citep{{Simons2002},
  {Tokunaga2002}} to the magnitudes in the JWST/NIRCam system using
the color transformations in Appendix~\ref{sec:color_conversion}.
Based on the pixel-to-pixel noise in the NIR images, the 10$\sigma$
detection limits are $m_{\rm F115W} =24.6$--24.8 mag, $m_{\rm F150W} =
24.2$--24.6 mag, and $m_{\rm F200W} = 23.6$--23.8 mag in the Vega
system.
The 10$\sigma$ limits vary slightly from region to region depending on
the exposure time and the presence of the Cloud 2 nebula.
The 10$\sigma$ limits in each region, as defined in
Section~\ref{sec:ident_cl}, are summarized in Table~\ref{tab:limit}.

\section{Results} \label{sec:Result}

\subsection{Identification of the Young Clusters} \label{sec:ident_cl}

\begin{figure}
  \begin{center}
    \gridline{\fig{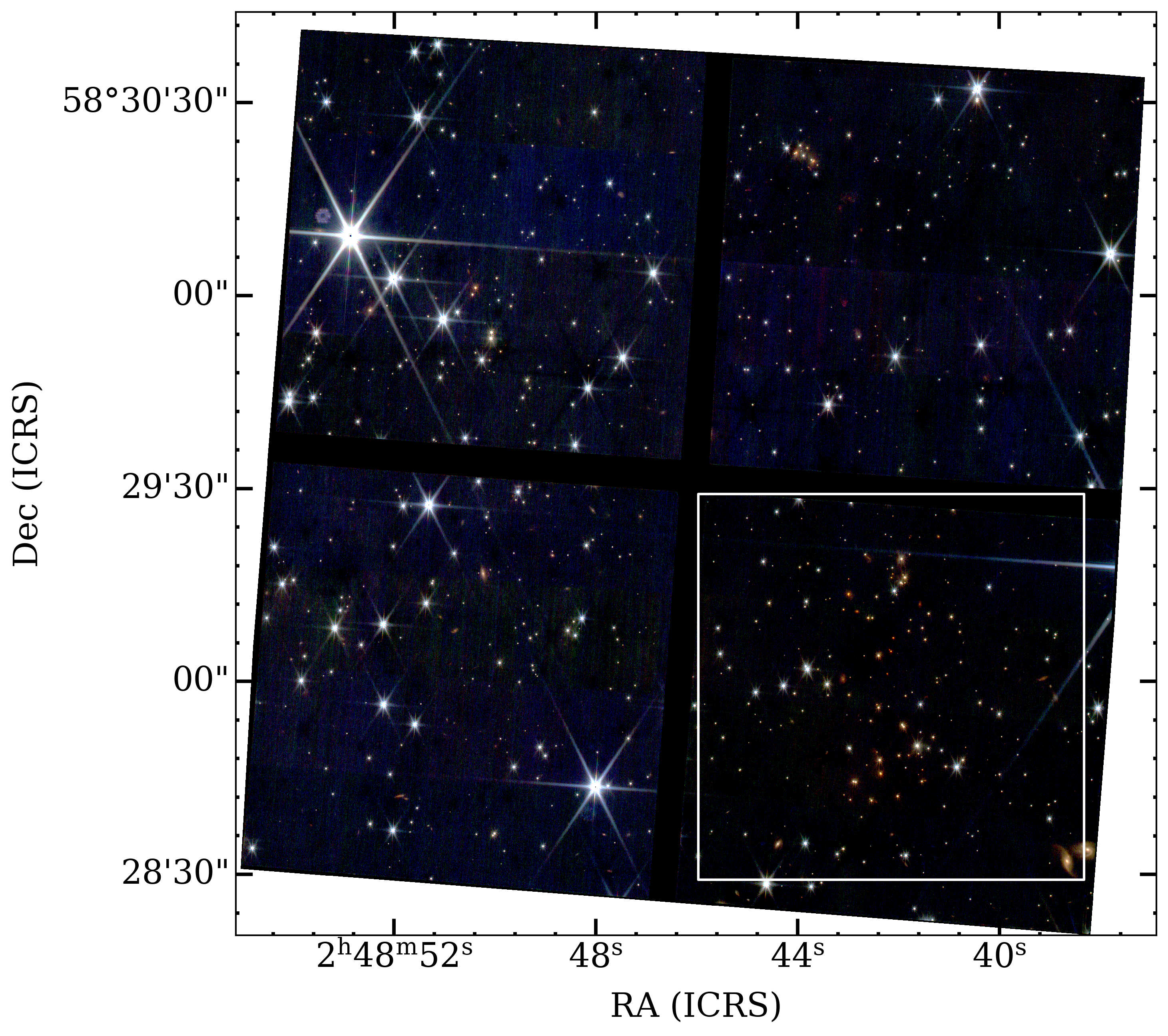}{0.95\textwidth}{}}
    \caption{Pseudocolor images of Cloud 2-N obtained with JWST/NIRCam
      produced by combining the NIR images, F115W (blue), F150W
      (green), and F200W (red).  The solid white square indicates the
      1 arcmin square centered on the Cloud 2-N cluster, shown in
      Figure~\ref{fig:CL2NScl}.}
\label{fig:3col_CL2N}
\end{center}
\end{figure}

\begin{figure}
  \begin{center}
    \gridline{\fig{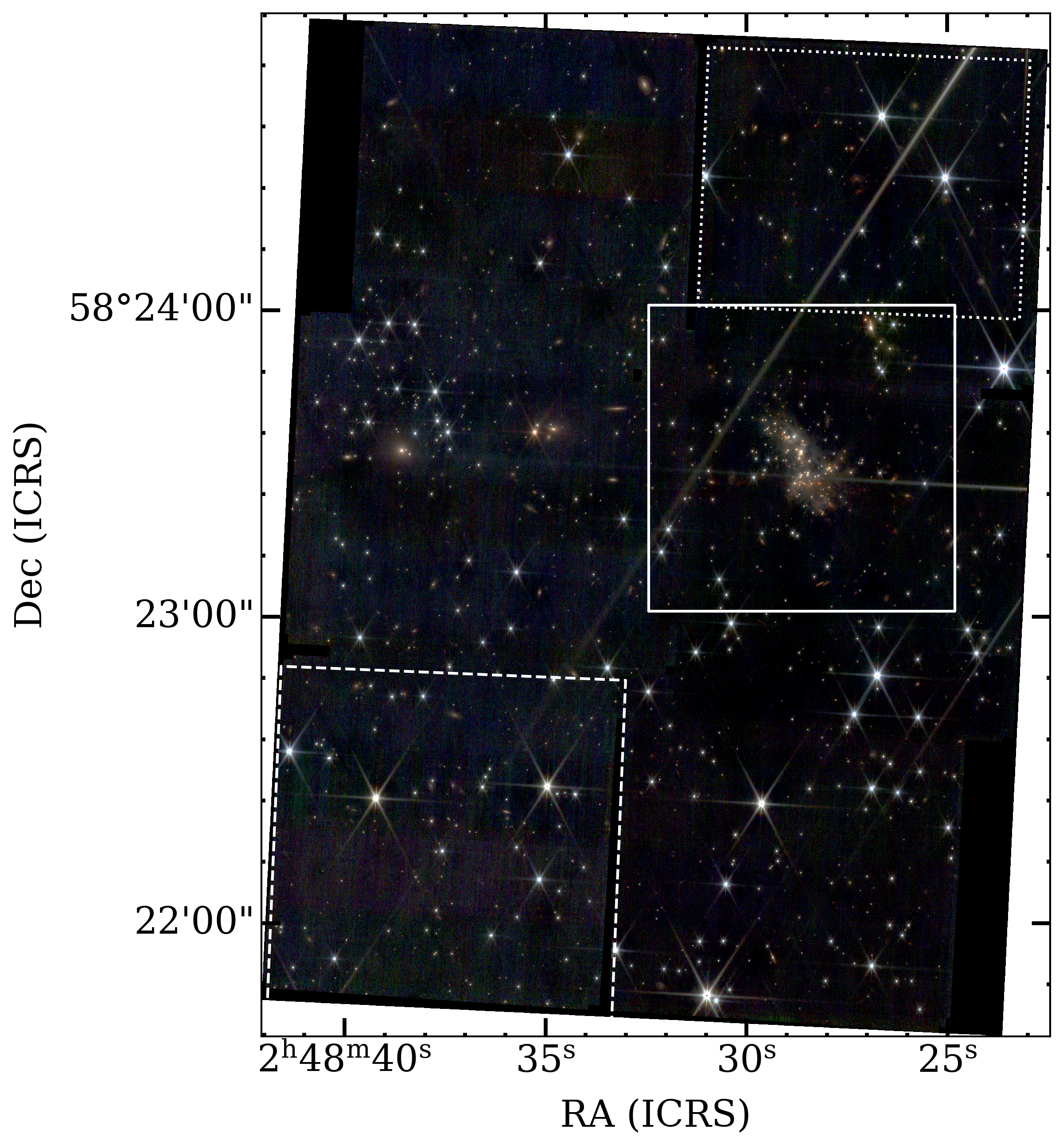}{0.95\textwidth}{}}
    \caption{Pseudocolor images of Cloud 2-S obtained with JWST/NIRCam
   produced by combining the NIR images, F115W (blue), F150W (green),
   and F200W (red).
   The solid white square indicates the 1 arcmin square centered on
   the Cloud 2-S cluster, shown in Figure~\ref{fig:CL2NScl}, while the
   white dashed square shows the control field.
   An addional control field is indicated by a dotted squre.}
\label{fig:3col_CL2S}
\end{center}
\end{figure}

In our previous studies with ground-based data, we discussed star
formation markers from NIR ($\le$2 $\mu$m) data only.
Here, in addition to the NIR data, we can use the longer wavelength
NIRCam and MIRI data to help identify regions of star forming activity
and individual protostellar sources.
Figures~\ref{fig:3col_CL2N} and ~\ref{fig:3col_CL2S} show the
pseudocolor images of Cloud 2-N and -S with JWST/NIRCam in the NIR.
In the MIR data, star-forming regions are more prominent than
in the NIR data.
Observations of a particularly dense region of the molecular cloud
confirmed that there are no star clusters with $N \ge 35$ (e.g.,
\citealt{LadaLada2003}) other than the Cloud 2-N, -S clusters, at
least within the field of view of the observations here.
An overview of star formation in Cloud 2, covered by JWST/NIRCam and
MIRI in this program, is discussed in \citet{Izumi2024}.
Because stars in such groups share the common heritage of being formed
more or less simultaneously from the same progenitor molecular cloud,
their observations are very well suited for deriving initial mass
functions under certain circumstances.

\begin{figure}
  \begin{center}
    \epsscale{0.5}
    \gridline{\fig{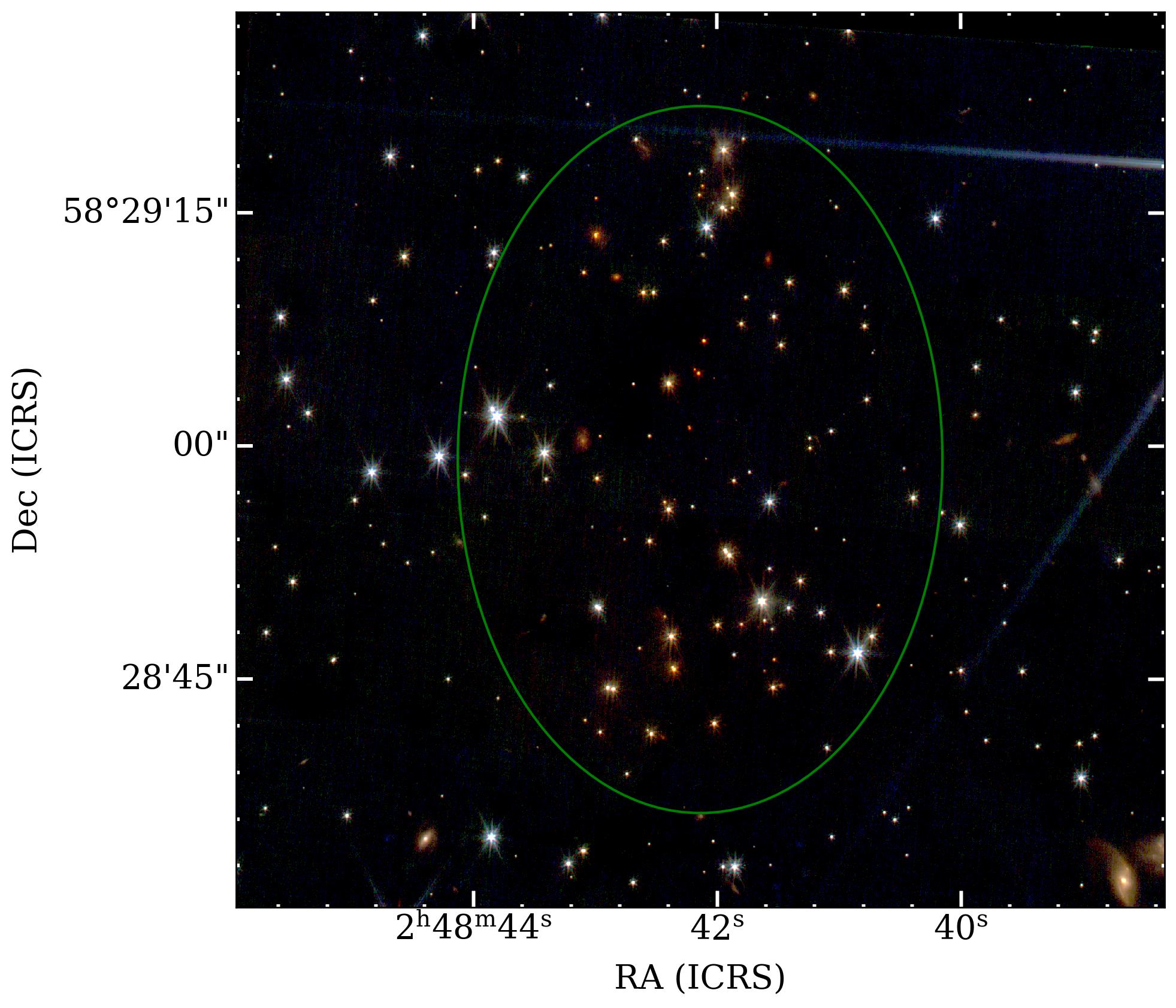}{0.5\textwidth}{}
              \fig{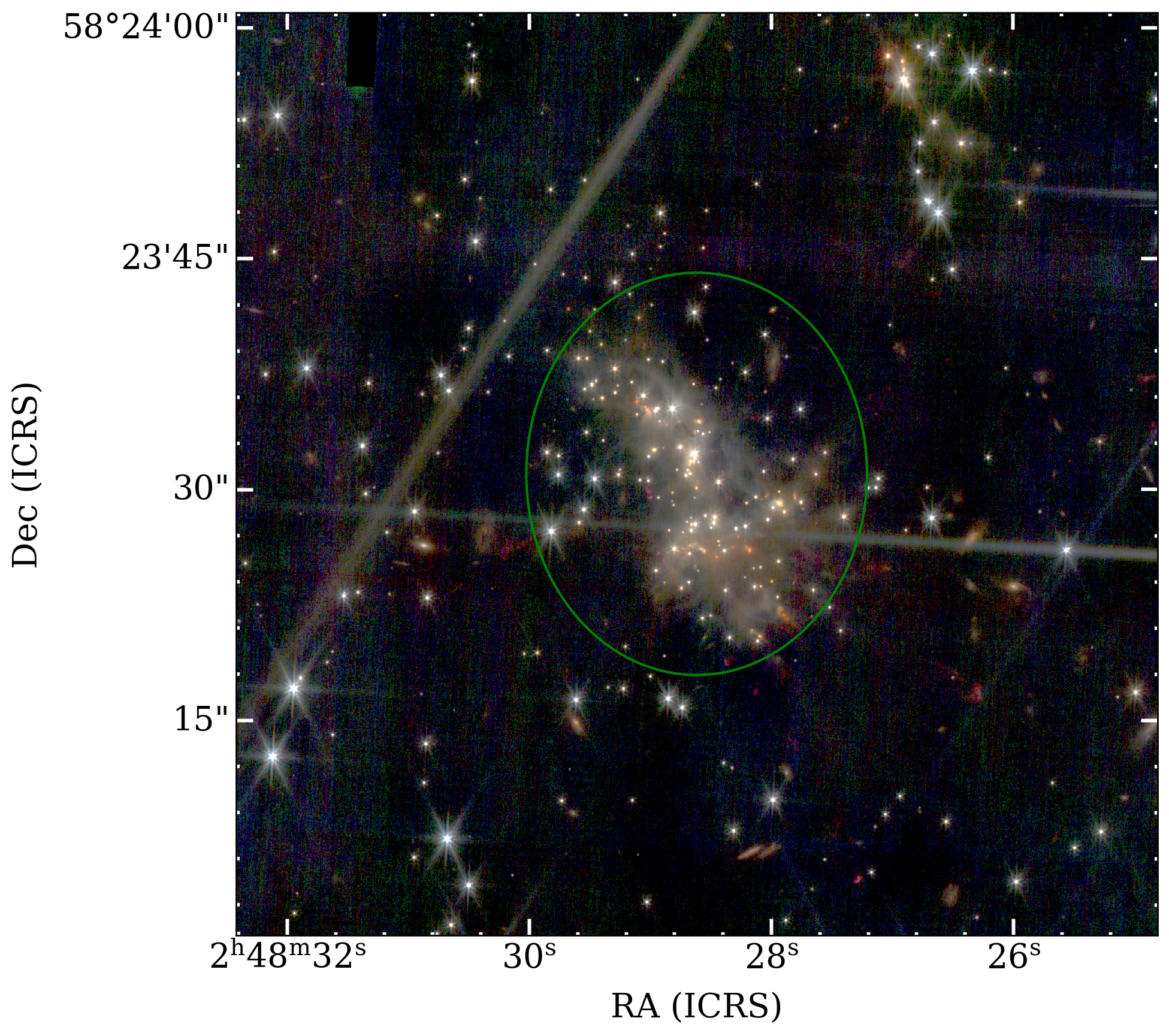}{0.48\textwidth}{}
              }
    \caption{Zoomed in F200W-band images of the Cloud 2-N and -S
      clusters. North is up and east is left.  The areas of the images
      are indicated by the white squares in
      figures~\ref{fig:3col_CL2N} and \ref{fig:3col_CL2S}, with the
      field of view of about $1' \times 1'$.  The positions of the
      cluster regions, which are identified in
      Section~\ref{sec:ident_cl} from the regions where particularly
      high stellar densities were detected in the MIRI images
      \citep{Izumi2024}, are shown as green ellipses.}
\label{fig:CL2NScl}
\end{center}
\end{figure}

Figure~\ref{fig:CL2NScl} shows NIR images of the Cloud 2 clusters
obtained in the NIRCam F200W band (left and right panels for the Cloud
2-N and -S clusters, respectively).
The regions where particularly high stellar densities were detected in
the MIRI images (see the bottom panels of Figures~5 and 6 in
\citealt{Izumi2024}) are indicated by ellipses and are defined as
cluster regions in this paper.
The ellipse for the Cloud 2-N cluster has a center coordinate
$(\alpha_{\rm 2000}, \delta_{\rm 2000}) = (02^{\rm h} 48^{\rm m}
42\fs1, +58^\circ 28' 59\farcs1)$, a minor radius of $r_\alpha =
13\arcsec$ in the direction of right ascension, a major radius of
$r_\delta = 16\arcsec$ in the direction of declination, while the
Cloud 2-S cluster has a center coordinate $(\alpha_{\rm 2000},
\delta_{\rm 2000}) = (02^{\rm h} 48^{\rm m} 28\fs6, +58^\circ 23'
31\farcs0)$,
a minor radius of $r_\alpha = 11\arcsec$ in the direction of right
ascension, a major radius of $r_\delta = 13\arcsec$ in the direction
of declination.
Note that the definitions of the cluster regions here are almost the
same as in our previous studies \citep{{Yasui2006}, {Yasui2008},
 {Yasui2009}}.

Cluster regions almost always contain foreground/background stars and
background galaxies.
Because Cloud 2 is located in the Outer Galaxy with $R_G \simeq
17$ kpc, the cluster regions occupy very small sizes on the sky
($\simeq$15$''$ radii) and the number of such contaminators is
expected to be very limited.
However, given the average distribution of stars along the line of
sight to this region of the Galaxy, we do expect a few non-cluster
sources to be within our defined cluster regions.
Therefore, it is necessary to subtract such contaminators by taking a
region on the sky that is close to, but otherwise unrelated to, Cloud
2 as a control field.
There is a 1 arcmin square region to the southeast of the Cloud 2-S
cluster in the NIRCam images that does not otherwise overlap the
star-forming regions judging from $^{12}$CO distributions (see
\citealt{Izumi2024} and Figure~11 in \citealt{Yasui2008}), WISE data
(Figure~2 in \citealt{Izumi2017}), and other multiwavelength data
\citep{Kobayashi2008}, shown as a white dashed square in
Figure~\ref{fig:3col_CL2S}.
We choose this area as our control field for evaluating the
contamination by non-cluster sources in later sections.

\subsection{Reddening properties} \label{sec:reddening}

There exist two types of reddening for sources in star-forming
clusters.
The first is extinction.
In particular, because the Cloud 2 clusters are located at such a
large distance ($D \simeq 10$ kpc) along the plane of the Galaxy,
members are subject to be extinguished by interstellar materials that
exist in the direction to the clusters.
In addition, clusters are embedded in molecular clouds, which also
cause significant extinction.
Note that the extinction from circumstellar materials is suggested
that the reddening law is approximately the same as that by
interstellar materials \citep{Bouvier2013}. 
Another reddening type is color excesses by circumstellar material. 
Circumstellar material is often present around young stars, which then
absorbs shorter wavelength radiation and reradiates it at longer
wavelengths (e.g., \citealt{Lada1992}).
In this section, we determine the distribution of reddening for each
of the objects in the Cloud 2 clusters and consider the distribution
as the probability distributions of reddening.  The two reddenings are
generated stochastically based on probability distribution functions
and reflect them in the pseudo-generated objects in when creating the
model luminosity function in Section~\ref{sec:model}.
For this purpose, we estimate the two reddenings for each source in Cloud 2
clusters and then derive their distribution for each cluster.

\begin{figure}
  \begin{center}
    \gridline{\fig{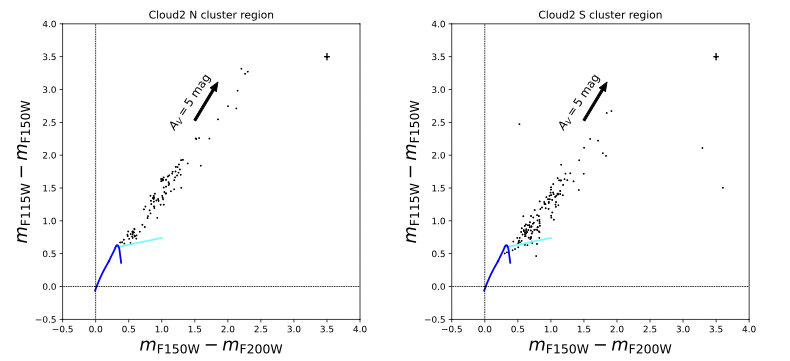}{0.9\textwidth}{}}
    \caption{($m_{\rm F150W} - m_{\rm F200W}$) vs. ($m_{\rm F115W} -
    m_{\rm F150W}$) color--color diagrams for the Cloud 2-N cluster
    region ({\it left}) and the 2-S region ({\it right}).  The blue
    curve in the lower left portion of each diagram is the locus of
    points corresponding to unreddened main-sequence stars.  The loci
    of classical T Tauri star (CTTS) are shown with cyan lines.  The
    black arrows show the reddening vectors for $A_V= 5$ mag.
    Typical uncertainty (1$\sigma$) of the colors are shown at the
    top-right corner.}
\label{fig:CC_Cloud2cl}
\end{center}
\end{figure}

We obtain reddenings for each source in Cloud 2 clusters using a
color--color diagrams based on \citet{Yasui2023}.
They used $H-K$ vs. $J-H$ diagram in NIR JHK-bands, but the ($m_{\rm
  F150W} - m_{\rm F200W}$) vs. ($m_{\rm F115W} - m_{\rm F150W}$)
diagram is used instead here, where $m_{\rm F115W}$, $m_{\rm F150W}$,
and $m_{\rm F200W}$ are magnitudes in the F115W, F150W, and F200W
bands, respectively, since they most closely reproduce the J, H, and K bands. 
The ($m_{\rm F150W} - m_{\rm F200W}$) vs. ($m_{\rm F115W} - m_{\rm
  F150W}$) color--color diagrams for the Cloud 2 clusters are shown in
Figure~\ref{fig:CC_Cloud2cl}.
All point sources that are located in the cluster regions and are
detected with more than 10$\sigma$ in all F115W, F150W, and F200W
bands are shown, 106 and 135 objects in total for the Cloud 2-N and -S
clusters, respectively.
We estimated the extinction ($A_V$) and the intrinsic (dereddened)
($m_{\rm F150W} - m_{\rm F200W}$) color [$(m_{\rm F150W} - m_{\rm
    F200W})_0$] for each star by dereddening it along the reddening
vector (shown with black arrows in Figure~\ref{fig:CC_Cloud2cl}) to
the young-star locus in the color--color diagram.
Because considering realistically possible IMFs, most of cluster
members are expected to be M0 type or later (M0 type stars with masses
of $\simeq$1 $M_\odot$), i.e., after the bend of the dwarf track
\citep[e.g.,][]{Bessell1988},
we approximated the young-star locus by extending the classical T
Tauri star (CTTS) locus (shown with cyan lines in
Figure~\ref{fig:CC_Cloud2cl}) for convenience,
and we used only those stars that are above the CTTS locus.
When each star is dereddened to the young-star locus on the
color--color diagram, the $A_V$ values are derived based on the
distance required to achieve the obtained amount of dereddening with
the reddening law while the intrinsic $(m_{\rm F150W} - m_{\rm
  F200W})_0$ are obtained from the value of $m_{\rm F150W} - m_{\rm
  F200W}$ on the young star locus.

Although the bandpasses of F115W, F150W, and F200W are reasonably
close to the JHK-band bandpasses in the conventional filter systems
(cf. generally, $J_{\rm eff} \simeq 1.25$ um, $H_{\rm eff} \simeq
1.65$ $\mu$m, and $K_{\rm eff} \simeq 2.2$ $\mu$m; see
\citealt{Stephens2004}), all of them have shorter wavelengths than the
corresponding JHK-band bandpass.
The filter bandpasses of F115W are F150W are shifted to the short
side by about 1/3 of the bandwidth compared to J and H bands,
while that of F200W is shifted to the short side by about half of the
bandwidth compared to K band.
This will lead to different basic properties of the color--color
diagram in the NIRCam filter system compared to those in the NIR JHK
filter systems.
As for the reddening law, there is a previous study using JWST/NIRCam,
where \citet{Wang2019} evaluated the relative extinction values
($A_\lambda / A_V$) in the JWST/NIRCam filter system, based on NIR
extinction law obtained from observations of red clump stars. 
For the CTTS locus, we derived the most plausible position of the CTTS
locus in the NIRCam filter system, shown as cyan lines (see details in
Appendix~\ref{sec:color_conversion}).
Supplementally, dwarf-star tracks in the NIRCam filter system are also
shown as blue curves (see detail in Appendix~\ref{sec:cc_nircam}).

\begin{figure}
 \begin{center}
   \gridline{\fig{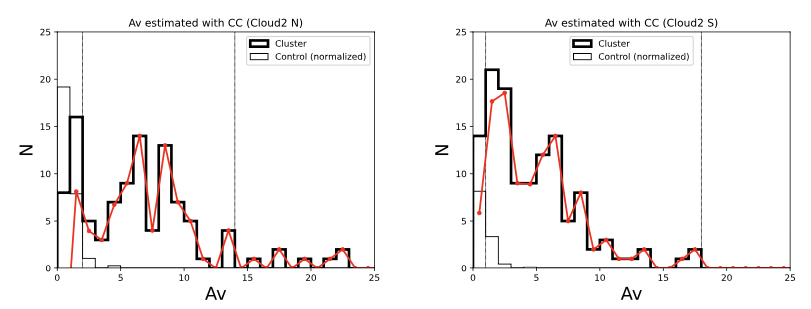}{0.9\textwidth}{}}
   \caption{$A_V$ distributions for stars in the regions of the Cloud
     2-N cluster (left panel) and of the 2-S cluster (right panel).
     The distributions for stars in the cluster regions are shown as
     thick lines, while those in the control field are shown as thin
     lines.  The distributions for stars in the control field are
     normalized by multiplying the ratio of the area of each cluster
     region to the area of the control field.  The distributions
     obtained by subtracting the normalized distribution for stars in
     the control field from the distribution for stars in the cluster
     regions are shown as red lines.  The vertical dotted lines show
     the $A_V$ ranges of the Cloud 2 clusters set in
     Section~\ref{sec:mass-av-sample} for defining the
     mass-$A_V$-limited sample.}
\label{fig:Av_CC}
 \end{center}
\end{figure}

\begin{figure}
 \begin{center}
   \gridline{\fig{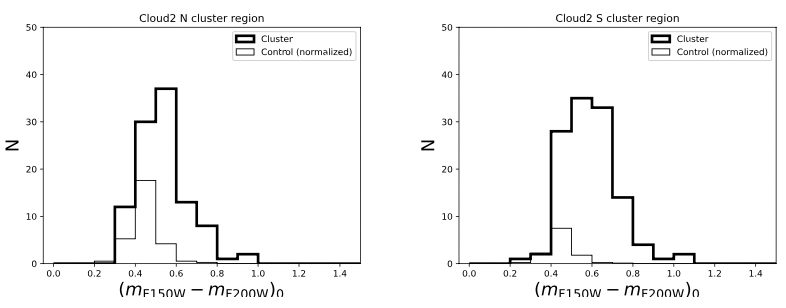}{0.9\textwidth}{}}
   \caption{Distributions of $(m_{\rm F115W} - m_{\rm F150W})_0$ for
     stars in the regions of the Cloud 2-N cluster (left panel) and of
     the Cloud 2-S cluster (right panel). The distributions for stars
     in the cluster regions are shown as thick lines, while those in
     the control field are shown as thin lines. The distributions for
     stars in the control field are normalized by multiplying the
     ratio of the area of each cluster region to the area of the
     control field.}
\label{fig:HK0_CC}
\end{center}
\end{figure}

\begin{figure}
  \begin{center}
    \gridline{\fig{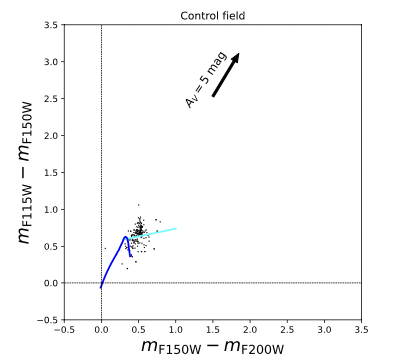}{0.45\textwidth}{}}
    \caption{Same figure as Figure~\ref{fig:CC_Cloud2cl}, but for the
      control field.}
\label{fig:colcol_CF}
\end{center}
\end{figure}

The resulting distributions of $A_V$ and $(m_{\rm F150W} - m_{\rm
  F200W})_0$ for the stars in the cluster regions are shown as thick
lines in Figures~\ref{fig:Av_CC} and \ref{fig:HK0_CC}, respectively.
To statistically remove the effects of sources that are located in the
cluster regions but are foreground or background stars, we also
derived the distributions in the control field using the color--color
diagram for stars in the control field, shown in
Figure~\ref{fig:colcol_CF}.
The obtained distributions were normalized by multiplying the ratio of
the area of each cluster region to the area of the control field and
are shown as thin lines in Figures~\ref{fig:Av_CC} and
\ref{fig:HK0_CC}.
In Figure~\ref{fig:Av_CC}, the distribution of stars in the cluster
region minus the distribution of stars in the control field is shown by
the red line, which is approximately the distribution of cluster
members.

In Figure~\ref{fig:Av_CC}, the $A_V$ distribution of stars in the
control field decreases monotonically with increasing $A_V$, reaching
almost zero at $A_V = 5$ mag.
In the cluster regions, on the other hand, the distribution extends to
larger values of $A_V \simeq 15$ mag for both clusters.
The reason why stars in the cluster region have such large $A_V$
values may be that they are subject to large extinction by the
interstellar medium due to the relatively large distance to Cloud 2.
In addition, the large dispersion of $A_V$ values should be due to the
fact that star-forming molecular clouds are still present in this
star-forming region.
Because \citet{KT2000} confirmed that no foreground clouds exist in
the direction of Cloud 2 from the $^{12}$CO survey data of the Five
College Radio Observatory (FCRAO) (Heyer et al. 1998), those in the
$A_V$ range whose counts in the cluster region are significantly
larger than those in the control field and those with large $A_V$
values can be considered cluster members.

The sources with $A_V \gtrsim 5$ mag clearly appear to be cluster
members, but there also appear to be sources with relatively small
$A_V$ of 1--2 mag for both Cloud 2-N and -S clusters. However, when
figures are produced in a different bin from Figure~\ref{fig:Av_CC}
(e.g., $\Delta A_V = 1.5$ mag), the peak is no longer seen in the
distribution for Cloud 2-N, while the peak is still seen for Cloud
2-S. This suggests that there are indeed cluster members with
relatively small values for Cloud 2-S only.
Therefore, most of stars in the range of $A_V \ge 2$ mag and $A_V \ge
1$ mag for Cloud 2-N and -S clusters, respectively, where the number
of objects in the cluster region is much larger than the number of
objects in the control region continuously, are considered to be from
Cloud 2 clusters.

It should be noted that all sources in the control field have
extinctions significantly lower than the median for objects in the
cluster area.
This may indicate that the background contamination is negligible, but
in fact it may simply indicate that while background stars are
present, the extinction due to molecular clouds is much lower in the
control field.
In the latter case, it would be incorrect to remove the influence of
background sources here.
To verify this, we set a region north of the Cloud 2-S cluster region,
shown as the region enclosed by a dotted square in
Figure~\ref{fig:3col_CL2S}, as another control field and obtained the
$A_V$ distribution in the same way as for the original control field.
In \citet{Izumi2024}, the region was confirmed to have no significant
star formaing activities and a CO molecular cloud column density
similar to the Cloud 2-N and -S cluster regions.
The derived $A_V$ distribution is quite consistent with the original
control field distribution.
Therefore, we can conclude that the influence of background stars in
these regions is negligible.

In Figure~\ref{fig:HK0_CC}, the $(m_{\rm F150W} - m_{\rm F200W})_0$
distribution of stars in the control field increases from $(m_{\rm
  F150W} - m_{\rm F200W})_0 = 0$ mag to 0.4--0.5 mag and then
decreases (to zero at about 0.6 mag),
while the distributions of stars in the Cloud 2-N and -S cluster
regions increase from $(m_{\rm F150W} - m_{\rm F200W})_0 = 0$ mag to
0.5--0.6 mag and extend up to $(m_{\rm F150W} - m_{\rm F200W})_0
\simeq 1.0$ mag.
The larger $(m_{\rm F150W} - m_{\rm F200W})_0$ of the stars in the
cluster region is due to the color excess produced by the
circumstellar material around the young stars. We assume that this
excess is only a F200W-band excess (\citealt{Strom1989}; abbreviated
``F200W-excess'').
We take the average value of the $(m_{\rm F150W} - m_{\rm F200W})_0$
of stars in the control field, which is estimated to be 0.45 mag, as a
typical $(m_{\rm F150W} - m_{\rm F200W})_0$ value of stars without any
surrounding materials, $(m_{\rm F150W} - m_{\rm F200W})_{0, {\rm
    crit}} = 0.45$ mag.

\begin{figure}[!h]
 \begin{center}
   \gridline{\fig{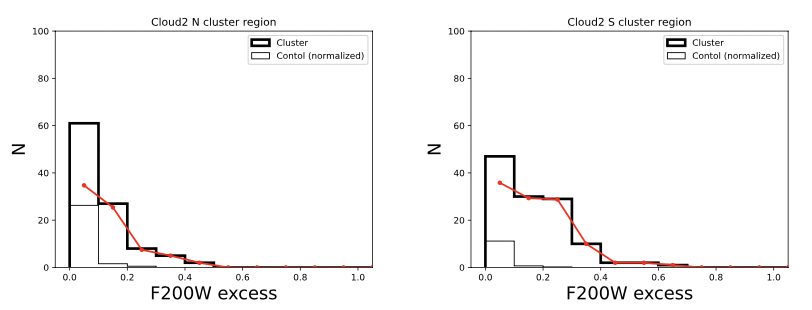}{0.9\textwidth}{}}
   \caption{F200W-excess distributions for stars in the regions of the
     Cloud 2-N cluster (left panel) and of the 2-S cluster (right
     panel).  The distributions for stars in the cluster regions are
     shown as thick lines, while those in the control field are shown
     as thin lines.  The distributions for stars in the control field
     are normalized by multiplying the ratio of the area of each
     cluster region to the area of the control field.  The
     distributions obtained by subtracting the normalized distribution
     for stars in the control field from the distribution for stars in
     each cluster region are shown as red lines.}
\label{fig:Kex_CC}
\end{center}
\end{figure}

We consider stars with $(m_{\rm F150W} - m_{\rm F200W})_0 \ge (m_{\rm
  F150W} - m_{\rm F200W})_{0, {\rm crit}}$ to be stars with a
F200W-excess, which we define as $(m_{\rm F150W} - m_{\rm F200W})_0$ minus
$(m_{\rm F150W} - m_{\rm F200W})_{0, {\rm crit}}$, while stars with 
$(m_{\rm F150W} - m_{\rm F200W})_{0} \le (m_{\rm F150W} - m_{\rm
  F200W})_{0, {\rm crit}}$ are considered to be stars without a
F200W-excess. The resulting distributions of F200W-excess are shown in
Figure~\ref{fig:Kex_CC}.
The distributions for stars in the Cloud 2-N and -S cluster regions are
shown in the left and right panels, respectively.
As in Figures~\ref{fig:Av_CC} and \ref{fig:HK0_CC}, distributions for
stars in the cluster regions are shown with thick lines, while those
for stars in the control field are shown with thin lines. The
subtracted distributions are shown with red lines. In a later section,
we use the subtracted distributions of stars for the Cloud 2-N and -S
clusters to construct model F200W-LFs for evaluating the IMF.

\subsection{Mass-$A_V$ limited samples} \label{sec:mass-av-sample}

\begin{figure}[!h]
\begin{center}
  \gridline{\fig{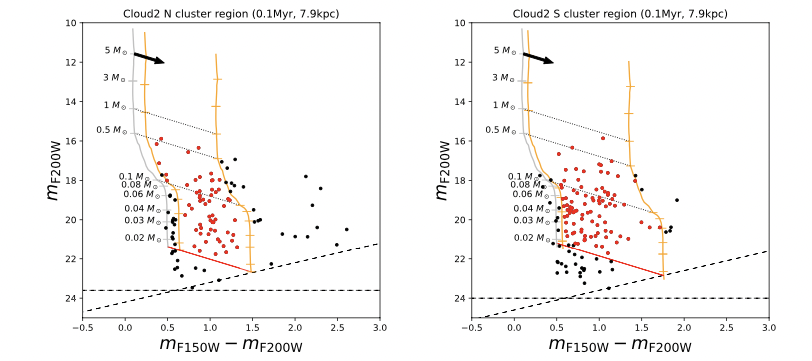}{0.9\textwidth}{}}
   \caption{$(m_{\rm F150W} - m_{\rm F200W})$ vs. $m_{\rm F200W}$
   color--magnitude diagram for the Cloud 2 clusters.  Only point
   sources located in the cluster region and are detected with more than
   10$\sigma$ in both the F150W and F200W bands are plotted.  The gray
   lines show isochrone models from \citet{Siess2000} for the mass range
   $3 < M/M_\odot \le 7$; and from \citet{D'Antona1997} for the mass
   range $0.017 \le M/M_\odot \le 3$.  A distance of 7.9 kpc and
   the age of 0.1 Myr are assumed.  The black arrows show the reddening
   vectors for $A_V=5$ mag from the isochrone models.  Stars located in
   the cluster region on the sky and are located between the orange
   lines on the color--magnitude diagram
   ($2 \le A_V \le 14$ mag and $1 \le A_V \le 18$ mag for the Cloud
   2-N and -S clusters, respectively),
   are identified as cluster members.  Identified cluster members are
   shown as red dots, while sources that are located in the cluster
   region but which are not considered to be cluster members shown as
   black dots.  The dashed lines show the 10$\sigma$ limits.
   The sensitivity limits are indicated by red lines.} 
\label{fig:CM_CL2NS}
\end{center}
\end{figure}

We extract mass-$A_V$-limited samples from the
color--magnitude diagram for each cluster in the same way as
\citet{Yasui2023}.
We constructed the $m_{\rm F150W} - m_{\rm F200W}$ versus $m_{\rm
  F200W}$
color-–magnitude diagrams for the point sources detected in the Cloud 2
cluster regions in Figure~\ref{fig:CM_CL2NS} (left and right panels
for the Cloud 2-N and -S clusters, respectively).
All point sources that are detected with more than 10$\sigma$ in both
the F150W and F200W bands are shown. The dashed lines mark 
the 10$\sigma$ limits. 
The gray lines in these figures show isochrone tracks for the age of
0.1 Myr:
from \citet{Siess2000} for the mass range $3 < M/M_\odot \le 7$; and
from \citet{{D'Antona1997}, {D'Antona1998}} for the mass range $0.017
\le M/M_\odot \le 3$. 
Model selection is discussed in Section~\ref{sec:model}.
The tickmarks on the isochrone models, which are shown in the same
colors as the isochrone tracks, correspond to the positions of stellar
masses 0.02, 0.03, 0.04, 0.06, 0.08, 0.1, 0.5, 1, 3, and 5
$M_\odot$. The arrows indicate the reddening vector for $A_V = 5$ mag.

We define the mass-$A_V$-limited sample for the  assumed distance and age.
We refer to Figure~\ref{fig:Av_CC} to limit the extinction range.
As discussed in Section~\ref{sec:reddening}, we set the minimum values
of $A_V$ to be 2 mag and 1 mag for the Cloud 2-N
and -S clusters, respectively.
On the larger $A_V$ side, more stars can be extracted in the cluster
field with a larger $A_V$ threshold, but
 if the $A_V$ range is set to a very large value, the mass detection
 limit becomes large when defining the Mass-$A_V$-limited sample.
Because in the cluster region the $A_V$ distributions for both
clusters break off around $A_V \simeq 15$ mag,
we set the upper limit of the $A_V$ range here as $A_V = 14$ and
$A_V = 18$ mag for the Cloud 2-N and 2-S clusters, respectively.
In this case, the mass-sensitivity limits in the defined $A_V$ range
are indicated by red lines, and the mass detection limits correspond
to $\simeq$0.02 $M_\odot$ for both clusters.

\begin{figure}[!h]
\begin{center}
  \gridline{\fig{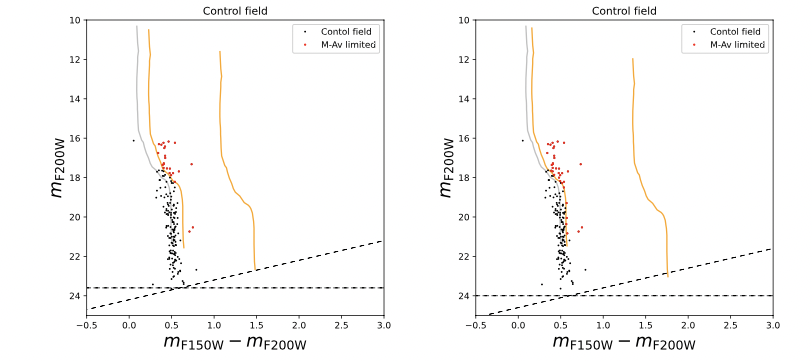}{0.9\textwidth}{}}
   \caption{Same figure as Figure~\ref{fig:CM_CL2NS}, but for the control field.
    The left and right panels are for the same control field, but just
    apply different $A_V$ thresholds for each cluster, the left and
    right panels ($2 \le A_V \le 14$ mag and $1 \le A_V \le 18$ mag)
    for the Cloud 2-N and -S clusters, respectively.} 
\label{fig:CM_control}
\end{center}
\end{figure}

In Figure~\ref{fig:CM_CL2NS}, our mass-$A_V$-limited samples are shown
in red and outliers are shown in black.
We note that the samples obtained here may still contain some
foreground (and sometimes background) sources; the color--magnitude
diagram of the control field is shown in Figure~\ref{fig:CM_control}
and shows that a few sources still meet our selection criteria.
In the control field, the most of objects are located around $(m_{\rm
  F150W} - m_{\rm F200W}) \simeq 0.5$ mag, whereas most {sources in
  the clusters} have $ (m_{\rm F150W} - m_{\rm F200W}) > 0.5$ mag,
with the center of distribution at $(m_{\rm F150W} - m_{\rm F200W})
\sim 1.0$ mag.
This is because the sources in the cluster are subject to local
extinction, considering the Cloud 2 clusters are embedded clusters.
Because of this feature, the number of stars in the control field
selected by the sample selection method is very small, especially at
$m_{\rm F200W} \gtrsim 18$ mag, where the PMS track is particularly
red. 
This suggests that the Mass-$A_V$-limited sample of clusters contains
very few foreground sources.
In fact, the effect of foreground stars on the F200W-LFs constructed
in Section~\ref{sec:clusterKLF} to derive the IMF is confirmed to be
negligible.

However, in order to assess and minimize the influence of foreground
stars,
we obtain the pseudo-mass-$A_V$-limited samples in the control field
in the same way as in the cluster region
(Figure~\ref{fig:CM_control}), and subtract the number of
pseudo-sources obtained in the control field from the number of
sources in each cluster region when constructing the F200W-LFs for the
IMF derivation in the next section (Section~\ref{sec:clusterKLF}).
In that section, the number of pseudo-mass-$A_V$-limited samples in
the control field is normalized by multiplying the ratio of the area
of each cluster region to the area of the control field.
The final number of objects in the Cloud 2-N and -S clusters are
estimated to be 52 and 91, respectively.

\subsection{Cluster F200W-LFs} \label{sec:clusterKLF}

\begin{figure}[!h]
  \begin{center}
    \gridline{\fig{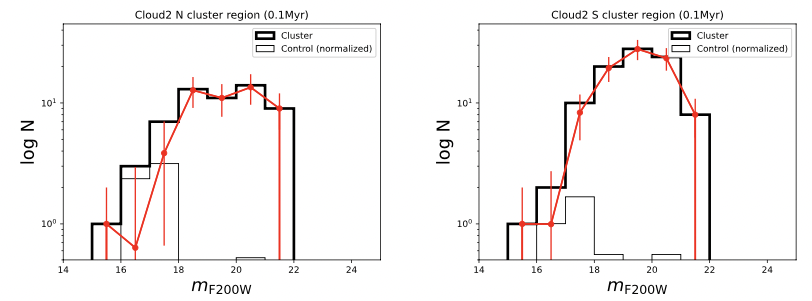}{0.9\textwidth}{}}
    \caption{The observed F200W-LFs for the Cloud 2 clusters (Cloud 2-N and -S
   clusters for the left and right panels, respectively).
   The F200W-LFs for sources in the cluster regions (the cluster
   region F200W-LFs) are shown with thick black lines, while those for
   sources in the control field (the control field F200W-LF) are shown
   with thin lines.
   The control field F200W-LF is normalized
   by multiplying the ratio of the area of each cluster region to the
   area of the control field. 
   The cluster F200W-LFs, obtained by subtracting the normalized
   counts from the control field F200W-LF from the counts for each
   cluster region F200W-LF, are shown as thick red lines.}
\label{fig:clKLFs}
\end{center}
\end{figure}

We constructed the F200W-LF for each cluster using the
mass-$A_V$-limited samples extracted in
Section~\ref{sec:mass-av-sample}.
(left and right panels in Figure~\ref{fig:clKLFs} for Clouds 2-N and
-S clusters, respectively).
In this figure, the F200W-LFs for sources in the cluster regions (the
cluster region F200W-LFs) are shown with thick black lines, while
those for sources in the control field (the control field F200W-LF)
are shown with thin lines.
The control field F200W-LF is normalized by multiplying the ratio of
the area of each cluster region to the area of the control field.
We subtracted the normalized counts from the control field F200W-LF
from the counts for each cluster region F200W-LF to obtain the cluster
F200W-LFs, which are shown as thick red lines. 
The F200W-LFs for the Cloud 2 clusters monotonically increase up to
$m_{\rm F200W} \sim 20$ mag for both clusters, and then decrease as
$m_{\rm F200W}$ magnitudes become even fainter.
In general, the F200W-LF peak reflects the IMF peak.  However, the
counts may just appear to decrease after $m_{\rm F200W} \sim 20$ mag
because we are taking mass-$A_V$-limited samples here, rather than
showing IMF turnover.  In fact, the color--magnitude diagrams
(Figure~\ref{fig:CM_CL2NS}) show that for $m_{\rm F200W} \lesssim 21$
mag there are stars in all $A_V$ ranges, sandwiched by two orange
lines (isochrone tracks with minimum and maximum $A_V$ values, i.e.,
all Av ranges), whereas for $m_{\rm F200W} \gtrsim 21$ mag there are
stars only in the larger $A_V$ ranges, sandwiched with red line
(showing the {sensitivity limit}) and a orange line (showing isochrone
track with maximum $A_V$ value).
Also, note that the counts of the control field are very small ($N
\lesssim 5$) in both clusters.
In particular, there are almost no counts of the control field around
the F200W-LF peak, suggesting that the influence of the foreground is
negligible.

Also, it should be noted that the detection completeness of stars with
$>$10$\sigma$ detection is generally almost one, whereas that of the
fainter stars is less than one (e.g., \citealt{Minowa2005}).
Actually, \citet{Yasui2008} confirmed the same results from estimation
of the detection completeness for the Subaru/MOIRCS ground-based data
by putting artificial stars at random positions in the Cloud 2 cluster
regions and checking whether they are detected in the same way as the
real objects.  Because data from JWST has much high qualities, e.g.,
high spatial resolutions and low background noise, this should be hold
(or the situation should be even better) for the JWST data here and
the brightness completeness may be at fainter magnitudes.  Considering
the 10$\sigma$ detection magnitude shown in Table~\ref{tab:limit}, the
completeness should be $\sim$1 in the all magnitude bins of the Cloud
2 clusters' F200W-LFs.

In later sections, the cluster F200W-LFs obtained here will be
compared to the model F200W-LFs that are constructed considering the
same mass-$A_V$-limited sample as for the observations, to derive the
IMF.
A similar method was used in the Trapezium cluster of
\citet{Muench2002}, but it was necessary to consider background
sources as contamination in addition to foreground sources for nearby
clusters such as the Trapezium cluster, thus accounting for complex
reddening.
On the other hand, in the case of Cloud 2 clusters, the effect of
foreground stars on the F200W-LF is confirmed to be almost
nonexistent, and there are no objects with large $A_V$ in the control
field, suggesting that the effect of background stars is also almost
nonexistent.
This is a similar trend for other star-forming clusters located in the
outer Galaxy compared to previous studies
\citep{{Yasui2016a}, {Yasui2021}, {Yasui2023}}.

\section{Derivation of the Cloud 2 clusters' IMF}

\subsection{Modeling}
\label{sec:model}

We derive the IMF for Cloud 2 clusters in the Outer Galaxy, which as
we have previously noted is in a low metallicity environment.
The most basic way to derive IMFs is to derive masses for individual
stars, and this method has often been used to derive IMFs for nearby
star forming regions (e.g.  \citealt{{Hillenbrand1997},
  {Luhman2000}}).
The most precise way to derive masses (and ages) for individual stars
to derive IMFs is using HR diagrams from spectroscopic observations. 
However, it takes a great deal of time to obtain sufficiently sensitive
spectra to be used for the classification when targeting these very
distant regions.
Conversely, here we derive the IMF from imaging data only, using a
method that was originally developed by \citet{Muench2002} to derive the
IMF for the Trapezium cluster in Orion and then modified by
\citet{Yasui2023}.
Note that the IMF obtained with this method for the Trapezium cluster
was consistent with the IMF obtained with the derived masses for
individual stars using spectroscopic data, but the method could derive
IMFs for much lower mass stars, compared to the IMFs obtained from
spectroscopic data (see \citealt{Muench2002}).

In our modeling, we first generated stars with masses determined
probabilistically according to IMFs. 
We assume a log-normal IMF \citep{Miller1979}, $dN / d \log m \propto
\exp (- C \times (\log m - \log M_C)$, and a tapered power-law (TPL)
IMF\footnote{\citet{De Marchi2010} defined the TPL IMF as $dN / dm
\propto m^{-\alpha} [1 - \exp(-m / M_C)]$, where $\alpha$ is the index
of the power law in the upper end of the mass function and $\beta$ is
the tapering exponent to describe the lower end of the IMF.
This formula can be rewitten as $dN / d \log m \propto m^{-\Gamma} \{1 - \exp
[- (m / M_C)^{\gamma + \Gamma}]\}$, 
where $\Gamma$ is the slope in the high-mass regime and $\gamma$
is the slope in the low-mass regime, and $\gamma + \Gamma$ is denoted
here as $\beta$.}
\citep{De Marchi2010}, $dN / d \log m \propto m^{-\Gamma} \{1 - \exp
      [- (m / M_C)^\beta]\}$.
The counts of F200W-LF first increase for fainter magnitudes ($m_{\rm
  F200W} \lesssim 18$ mag), then become almost flat ($m_{\rm F200W}
\simeq 18$--20 mag), before decreasing ($m_{\rm F200W} \gtrsim 21$
mag),
but it is not obvious at this stage whether the trend in F200W-LFs
counts reflects a similar trend in IMFs
(Section~\ref{sec:clusterKLF}).
Even these cases can expressed by log-normal or TPL IMF: in the case
of monotonic increase, the peak mass should be estimated to be smaller
than the minimum mass range, and in the case of flattening after
monotonic increase, the peak mass should be estimated to be slightly
larger than the minimum mass range.
The ranges of parameters for both IMF indices are set within
relatively realistic ranges of values: $C=0.1$ to 10 and $\log M_C =
-2.0$ to 0.0 for log-normal IMF, and $\Gamma = 0.5$ to 2.0, $\log M_C
= -2.0$ to 0.0, and $\beta = 0.5$ to 4.0 for TPL IMF.
All parameter combinations are tried in the set parameter range, in
increments of 0.1 for all parameters.
The IMF mass range is set to cover all possible ranges obtained from
mass-$A_V$-limited samples 
the magnitude range of F200W LF used for model fitting discussed in
Section~\ref{sec:derivedIMF}.

We then determined NIR luminosities for the artificially generated
sources based on the IMFs.
We used the mass–luminosity (M--L) relations employed in
Section~\ref{sec:mass-av-sample} for the isochrone tracks in the
color–magnitude diagram. The M--L relation assuming the most reliable
distance (7.9 kpc) and age (0.1 Myr) is shown as a black line in
Figure~\ref{fig:MLs}.
Regarding the choice of the isochrone model, \citet{{Muench2000},
  {Muench2002}} found that variations in the pre-main-sequence (PMS)
M–L relation, which result from differences in the adopted PMS tracks,
produce only small effects in the form of the model luminosity
functions, and these effects are mostly likely not detectable
observationally.
\citet{LadaLada2003} also pointed out that the predicted luminosities
are essentially degenerate with respect to different models (see their
Figure 7), and they suggested that this should be due to the fact that
the luminosity of a PMS star is determined by very basic physics,
simply the conversion of gravitational potential energy to radiant
luminosity during Kelvin-Helmholtz contraction.

\citet{Yasui2023} compared low-metallicity models (down to $[{\rm
    M/H}] = -2.0$) with solar metallicity models and found that the
differences in the M-L relation due to changes in metallicity within
the same group of models were very small. They also confirmed that the
differences in metallicity were significantly smaller than the
differences between models by different groups. Therefore, the model
with solar metallicity ($[{\rm M/H}] = 0.0$) was adopted here.
We also adopt the K-band luminosity for use as the F200W-band luminosity.
The difference between these luminosities is discussed in
Section~\ref{sec:uncertainties}.
The isochrone model for the age of the Cloud 2 clusters, 0.1 Myr, was
used, and age spread was not taken into account here because
\citet{Muench2000} found that variations in the cluster-age spread
have only a small effect on the form of the KLF and would also be
difficult to distinguish observationally.

The effects of M--L relations by the difference of filter system will
be discussed in Section~\ref{sec:uncertainties}.
After conversion of mass to luminosity, luminosity was converted to
NIR magnitudes considering the distance, then the effects of
reddening are also considered.
The reddening (extinction and infrared disk excess) was generated
stochastically using the reddening probability distribution for the
Cloud 2 clusters obtained in Section~\ref{sec:reddening}.

Obtained magnitudes and colors for pseudo-generated sources are
confirmed whether they satisfy the criteria for mass-$A_V$-limited
sample in Section~\ref{sec:mass-av-sample}, and if they do not meet
the criteria, they are not counted in the model
F200W-LF.
This process was repeated until the number of stars satisfying this
condition became equal to the number of stars in the
mass-$A_V$-limited samples for the Cloud 2 clusters that we had
obtained by subtracting the number of samples in the control field
from the number of samples in the cluster region.
For each synthetic cluster, the obtained magnitudes were binned by one
magnitude and compared to the F200W-LFs of the observed clusters to
derive the best IMF.
For both Cloud 2-N and -S, 100 independent F200W-LFs were generated
for each IMF. The average and corresponding 1$\sigma$ standard
deviation of these 100 F200W-LFs were calculated.

\subsection{The IMFs Derived for the Cloud 2 clusters}
\label{sec:derivedIMF}

For deriving best-fit IMF for the Cloud 2 clusters, chi-square test
are carried out.
It should be noted that there must be sufficiently large numbers of
frequencies in the bins for this test.
According to a textbook \citep{Wall2012}, the expected frequencies
should be $>$5 in $>$80\% of the bins, in particular because of the
severe instability at $<$5 counts per bin.  To meet this requirement,
four bins of $m_{\rm F200W} = 18$--22 mag for Cloud 2-N and {five
  bins of $m_{\rm F200W} = 17$--22 mag} for Cloud 2-S can be used
for the fit.
Because the log-normal and TPL IMFs have two and three parameters,
respectively, only log-normal IMF can be used for the Cloud 2-N while
both log-normal and TPL IMFs can be used for the Cloud 2-S.
As a result of F200W-LF fitting, we obtained reduced $\chi^2$ values
of $\simeq$1 from the best-fit IMF parameter sets for both clusters.
The obtained best-fit IMF parameters for the Cloud 2-N and -S clusters
are shown in Tables~\ref{tab:KLFfit_LN} and \ref{tab:KLFfit_TPL} for
log-normal and TPL IMFs, respectively, with the ranges corresponding
to the 68\% confidence levels shown in brackets.
The model F200W-LFs for the Cloud 2-N and -S clusters using the
parameters for the best-fit log-normal IMFs are shown in
Figure~\ref{fig:fitKLF}, along with observed F200W-LFs (left and right
panels for Cloud 2-N and -S clusters, respectively).
The red lines show best-fit model F200W-LFs with 1$\sigma$ standard
deviation, while blue histograms show observed cluster F200W-LFs.  The
bins of the fit range are indicated by filled blue squares, and the
others by open blue squares.
The best-fit IMFs for the Cloud 2-N and -S clusters are also shown in
Figures~\ref{fig:fitIMF_N} and \ref{fig:fitIMF_S}, respectively, shown
as black lines.
For reference, IMFs obtained in previous studies in the solar
neighborhood are also shown as colored lines:
\citet[magenta]{Salpeter1955}, \citet[green]{Scalo1998},
\citet[red]{Muench2002}, \citet[blue]{Miller1979}, and
\citet[orange]{Kroupa2002}.
All of the IMFs are normalized to 0 on the vertical axis at a mass of
1 $M_\odot$.

The $M_C$ values are estimated as $\log M_C = -1.5^{+0.1}_{-0.2}$ for
the Cloud 2-N cluster based on log-normal IMF, and $\log M_C =
-1.6^{+0.2}_{-0.4}$ and $-1.4^{+0.4}_{-0.3}$ for the Cloud 2-S
clusters based on log-normal IMF and TPL IMF, respectively.
Because the mass detection limits are 0.02 $M_\odot$ ($-$1.70 on the
log scale) for both Cloud 2 clusters, which are lower than the $M_C$
value obtained here, our observations are considered to cover the IMF
peaks for both clusters and the $M_C$ values obtained here are
considered to be significant.
This is the first derivations of $M_C$ for star-forming clusters in a
low-metallicity environment, which was achieved due to the quite high
sensitivities of the JWST.
We note that the $M_C$ values for both clusters are very similar
although there is some scatter in each value.

In the fitting results assuming log-normal IMF for the Cloud 2-S
cluster, the high-mass slope appears unrealistically steep although
they cannot be rejected from the chi-square test.
This may be because the log-normal function defines the IMF
symmetrically, and the slope on the high-mass side and the low-mass
side are only defined at the same value.
Therefore, the fit may not work very well for the Cloud 2-S cluster,
where the fit was performed over a relatively wide range of grades
(i.e., corresponding to a wide mass range).
The TPL IMF seems to improve on this point although the uncertainties
here is very large.
In any case, the derivation of $M_C$, the most noteworthy point of
this paper, yielded similar values for both assumed IMFs within the
error range.

Here the F200W-LF bins are taken as $m_{\rm F200W} = 15.0$--16.0,
16.0--17.0, ..., and 21.0--22.0 mag. The fits of LFs with the center
of the bin shifted by 0.5 mag ($m_{\rm F200W} = 15.5$--16.5,
16.5--17.5,...) are also attempted.
As a result, the obtained best-fit IMF parameters assuming log-normal
IMF for the Cloud 2-N are
$\log M_C = -1.5^{+0.3}_{-0.5}$ ($C=0.9^{+2.9}_{-0.8}$), while the
parameters assuming log-normal and TPL IMFs for the Cloud 2-S cluster
are
$\log M_C = -1.7^{+0.2}_{-0.3}$ ($C=2.2^{+1.2}_{-0.9}$) and
$\log M_C = -1.7^{+1.5}_{-0.3}$ ($\Gamma=0.9^{+1.1}_{-0.2}$ and 
$\beta=0.9^{+2.4}_{-0.3}$), respectively.  
The obtained $M_C$ values are not significantly different from those
obtained with the original LF bin settings.

We also note that there are large uncertainties in the derived IMF
parameters, especially for parameters related to high-mass slope, $C$
for log-normal IMF and $\Gamma$ for TPL IMF.
The main reason for this should be simply the small number of members
in the two clusters (52 and 91 for the Cloud 2-N and -S clusters,
respectively), as discussed in \citet{Yasui2023}.
In particular, the $\Gamma$ in the TPL IMF fit for the Cloud 2-S
cluster shows that the 68 \% confidence level is satisfied for almost
all the originally input parameter ranges, suggesting that the
uncertainties of $M_C$ be even larger. However, since fitting with
parameters that are far from reality is probably meaningless, further
fitting with a wider range of parameters is not performed here.


\section{Discussion} \label{sec:Discussion}

\subsection{Uncertainties in the derived $M_C$ for Cloud 2 clusters}
\label{sec:uncertainties}

\begin{figure}
\begin{center}
  \gridline{\fig{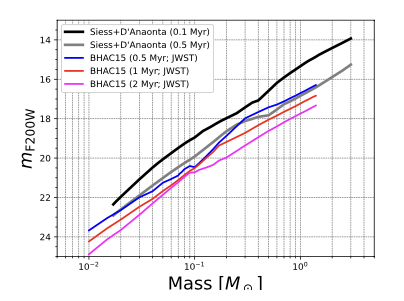}{0.5\textwidth}{}}
  \caption{Comparison of theoretical mass-luminosity ($m_{\rm F200W}$)
    relations.
    The M--L relation that is used in the modeling in
    Section~\ref{sec:model} \citep{{Siess2000}, {D'Antona1997},
      {D'Antona1998}} for the age of 0.1 Myr is shown with a black
    line, while the relation by the same reference but for the age of
    0.5 Myr is shown with a gray line.
    The M--L relations in the JWST/NIRCam filter system by
    \citet{Baraffe2015} for the ages of 0.5, 1, and 2 Myr are shown
    with blue, red, and magenta lines, respectively.
    The distance of $D=7.9$ kpc is assumed.}
\label{fig:MLs}
\end{center}
\end{figure}

\begin{figure}[!h]
\begin{center}
  \gridline{\fig{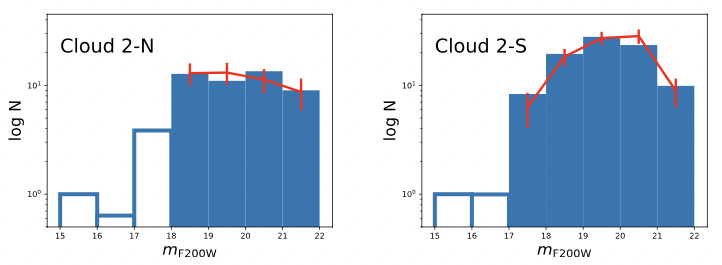}{0.9\textwidth}{}}
  \caption{Model F200W-LF with the parameters of the best-fit IMF for
    the Cloud 2-N and -S clusters.
    The best-fit model F200W-LF is shown as the red line with
    1$\sigma$ standard deviation, which are from 100 trials for the
    best-fit log-normal/TPL IMF parameter sets for the Cloud 2-N/-S
    clusters, while the observed cluster KLF for the age is shown as
    the blue histogram.
    The bins of the fit range are indicated by filled blue squares,
    and the others by open blue squares.}
\label{fig:fitKLF}
\end{center}
\end{figure}

\begin{figure}[!h]
  \begin{center}
    \gridline{\fig{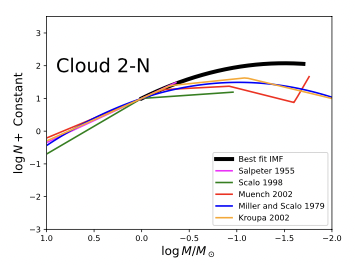}{0.45\textwidth}{}}
    \caption{The best-fit IMF for the Cloud 2-N cluster.
    The best-fit IMF assuming a log-normal IMF is shown as the black
    line.  The cluster IMF is also compared to IMFs previously
    obtained in the field and in nearby star clusters:
    \citet[magenta]{Salpeter1955}, \citet[green]{Scalo1998},
    \citet[red]{Muench2002}, \citet[blue]{Miller1979}, and
    \citet[orange]{Kroupa2002}.
    All of the IMFs are normalized to 0 on the vertical axis at a mass
    of 1 $M_\odot$.}
\label{fig:fitIMF_N}
\end{center}
\end{figure}

\begin{figure}[!h]
  \begin{center}
    \gridline{\fig{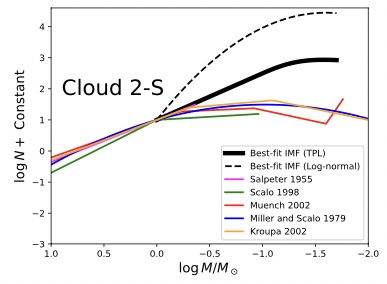}{0.45\textwidth}{}}
   \caption{The best-fit IMF for the Cloud 2-S cluster.  The
     best-fit IMFs assuming the TPL/log-normal IMFs are shown as the
     black thick/dashed lines.
     The cluster IMF is also compared to IMFs previously obtained in
     the field and in nearby star clusters:
     \citet[magenta]{Salpeter1955}, \citet[green]{Scalo1998},
     \citet[red]{Muench2002}, \citet[blue]{Miller1979}, and
     \citet[orange]{Kroupa2002}.
     All of the IMFs are normalized to 0 on the vertical axis at a
     mass of 1 $M_\odot$.}
\label{fig:fitIMF_S}
\end{center}
\end{figure}

Very similar values of $M_C$ are obtained for Cloud 2-N and -S
clusters, and although the $M_C$ values have large uncertainties, this
suggests that the value is typical for the region, $\log M_C \simeq
-1.5$.
We discuss possible uncertainties of $M_C$ besides
the uncertainties obtained from the F200W-LF fitting in
Section~\ref{sec:derivedIMF}.
First, as for the distance of the Cloud 2 clusters, the astrometric
distance $D=7.9$ kpc from Gaia DR3 was adopted in this paper
(Section~\ref{subsec:target}).
The distance has uncertainties of $\simeq$1 kpc, which cause a
difference of distance modulus $\Delta m = 0.3$ mag.
The $M_C$ obtained from the F200W-LF fit assuming the most reliable
age (0.1 Myr), $\log M_C \simeq -1.5$ or $M_C = 0.03$ $M_\odot$,
corresponds to $m_{\rm F200W}= 21.4$ mag under $A_V = 5$ mag.
In the case of smaller (larger) distances by 1 kpc, the magnitude
corresponds to a decrease (an increase) in mass of 0.01 $M_\odot$ (see
the black line in Figure~\ref{fig:MLs}).

It is possible that some objects in this study cannot be
  resolved at the spatial resolution of $\lesssim$500 au
(Section~\ref{subsec:target}); this will influence the derived $M_C$
value.
The binary fraction is known to be smaller for lower-mass stars
\citep{Offner2023}. 
In the mass range around $M_C$, the fraction is $\simeq$20\% at 0.1
$M_\odot$ and $\simeq$10\% at 0.05 $M_\odot$.
For FGK-type dwarfs, the fraction of close binaries ($<$10 au) is
known to increase strongly with decreasing metallicity, while the
fraction of wide binaries ($\gtrsim$200 au) is known to be relatively
constant over $-1.5 \lesssim {\rm [Fe/H]} \le 0.5$ \citep{Moe2019}.
If this is also the case for the low mass stars such as those targeted
here, then there may be more binaries in Cloud 2, which is in a
low-metallicity environment.
The binary fractions for open clusters are known to be consistent with
the fractions listed above, which are mainly derived from field stars,
for close and intermediate binaries, while those for wide binaries are
smaller than the field values.
The discrepancies in the wide binary fraction exist between the dense
Orion Nebula Cluster and the sparse Taurus star-forming region, which
exhibit a deficit and excess, respectively, compared to the field main
sequence stars \citep{{Duchene2018},{Kraus2011}}. 
The two clusters in Cloud 2 have different stellar densities (see
Figure~8 in \citealt{Yasui2008}):
the Cloud 2-N cluster is a Taurus-like sparse cluster ($\simeq$20
stars pc$^{-2}$), while the Cloud 2-S cluster is Orion Nebula Cluster
(ONC)-like dense cluster ($\simeq$70 stars pc$^{-2}$).
Despite these differences, the derived $M_C$ values for the clusters
are nearly identical, implying that this effect is very small, if it
exists at all.
Taking into account the unresolved binaries, the stellar mass and thus
the $M_C$ is estimated to be larger than it actually is.
We assume the extreme case where all objects are unresolved binaries
and the binaries have the same mass.
The flux corresponding to $M_C$ ($m_{\rm F200W} =21.4$ mag) would
consist of two objects with $m_{\rm F200W} =22.2$ mag, corresponding
to a mass reduction of 0.01 $M_\odot$ (see the black line in
Figure~\ref{fig:MLs}).

In \citet{Chabrier2003}, it was argued that the discrepancy between
the $M_C$ of the IMF for the unresolved system IMF ($\sim$0.2
$M_\odot$) and that for the single-star IMF ($\sim$0.08 $M_\odot$) is
explained by considering unresolved stars.
Since also for the target clusters here the sources cannot be resolved
within $\lesssim$500 au, the obtained $M_C$ should be the upper limit
of the single-star IMF, and the single-star IMF is expected to have an
even smaller value than the $M_C$ here.
As mentioned above, it is suggested that the overall binary fraction
tends to increase with decreasing metallicity, and taking this effect
into account, the single-star IMF may take a value even smaller than
that expected from the ratio of the system IMF to the single-star IMF
in the solar neighborhood.

In the F200W-LF fitting, we used K-band luminosities from the M--L
relation by \citet{{D'Antona1997}, {D'Antona1998}} for low-mass stars,
$0.017 \le M/M_\odot \le 3$.
Although the effects of adopting different models are thought to be
mostly likely not detectable observationally as discussed in
Section~\ref{sec:model},
we checked the effects of differences in filter systems (F200W vs. K)
adopted in the models.
The model by \citet{{D'Antona1997}, {D'Antona1998}} consider canonical
NIR filter system, while the model by \citet{Baraffe2015} provides
optical/NIR/MIR magnitudes in a variety of filter systems, including
JWST/NIRCam filter system.
The model by \citet{Baraffe2015} covers mass range of 0.01--1.4
$M_\odot$.
Because only models for ages of $\ge$0.5 Myr (older than clusters'
ages) are available, we could not use it for the F200W-LF fitting in
Section~\ref{sec:model}, but here we compare the two models for an age
0.5 Myr.
In Figure~\ref{fig:MLs}, the M--L relation by \citet{Baraffe2015} for
the age of 0.5 Myr in the JWST/NIRCam is shown with a blue line, while
that used in the modeling in Section~\ref{sec:model}
\citep{{Siess2000}, {D'Antona1997}, {D'Antona1998}} for the same age
is shown with a gray line. 
In the mass range where both models exist, the difference in
magnitudes between the two models at the same mass is 0.2 mag on
average (0.4 mag at most), which is significantly smaller than the
F200W-LF bin of 1 mag.
In the magnitude corresponding to the mass of $M_C$ ($m_{\rm F200W}
=21.4$ mag), the mass estimated from \citet{Baraffe2015} is almost
identical to that from \citet{{D'Antona1997}, {D'Antona1998}}.

The final concern is the uncertainty in the cluster ages. 
Although the discussion proceeded by assuming the most plausible age of
the clusters, $\simeq$0.1 Myr (Section~\ref{subsec:target}), the
F200W-LF fitting for older ages are also attempted.
It is known that the embedded phases of clusters last about 2--3 Myr
\citep{LadaLada2003}, and indeed clusters are known to be sufficiently
embedded in molecular clouds \citep{Izumi2024} to suggest that the age
of clusters should not exceed 2 Myr at most.
We performed F200W-LF fitting for ages up to 2 Myr (0.5, 1, and 2 Myr)
as well for the both Cloud 2 clusters, using the M-L relations in the
JWST filter system discussed above.
The $M_C$ is estimated as $\log M_C = -0.9$ to $-$1.0 for all assumed
ages for both Cloud 2 clusters.
The estimated value is also approximately consistent with the value
read from Figure~\ref{fig:MLs}, which in turn supports the fits here
work well.

In summary, it is suggested that the $M_C$ can range from {0.02--0.1
  $M_\odot$ ($\log M_C \simeq -1.7$ to $-$0.9)} although the $M_C$
obtained here have various kind of uncertainties.
In the next section, we discuss the possible metallicity dependence of
$M_C$ by comparison with $M_C$ for other derived regions in previous
studies.
While it should be noted that the uncertainty of the $M_C$ in Cloud 2
is somewhat large, the $M_C$ here is found to be smaller compared to
previous derivations.

\subsection{Possible metallicity dependence of $M_C$}

Because star-forming open clusters generally have similar
characteristics, e.g., young age ($\sim$$10^6$--$10^7$ yr) and
moderate cluster scale ($N_* \lesssim 10^3$) \citep{Portegies
  Zwart2010}, they are a good basis for comparison to evaluate the
metallicity dependence of $M_C$.
Previous studies compiled $M_C$ for local open clusters, and found
$M_C$ in all regions is about constant, $\log M_C /M_\odot \sim 0.5
\pm 0.5$, or $M_C \sim 0.3$ $M_\odot$ \citep[e.g.,][]{Elmegreen2008}. 
\citet{Elmegreen2008} suggested that the thermal Jeans mass ($M_J$),
which is expected to be proportional to $M_C$, depends weakly on
environment, such as density, temperature, metallicity, and radiation
field.
The $M_C$ values for the Cloud 2 clusters we have derived here {($M_C
  \simeq -1.5$)} are lower than the $M_C$ values estimated for local
open clusters although with larger uncertainties ($\simeq$0.5 in log
scale).
The lower $M_C$ values of the Cloud 2 clusters suggest that $M_C$ is
correlated with metallicity.

Because globular clusters have low metallicities ($[{\rm Fe/H}]
\simeq -0.70$ to $-$2.2) similar to Cloud 2, but have much older ages
($\sim$$10^{10}$ yr old), they are also good comparisons for understanding
the evolution of the IMF.
\citet{Elmegreen2008} and \citet{Paresce2000} claim that the $M_C$ of
the global clusters is approximately similar to other local regions,
while \citet{De Marchi2010} claimed that $M_C$ for globular clusters
are higher ($M_C \simeq 0.33$ $M_\odot$) than young open clusters
($M_C \simeq 0.15$ $M_\odot$).
Despite the fact that the Cloud 2 clusters have a similarly low
metallicity (although on the higher end of metallicity among globular
clusters), the obtained $M_C$ values for the Cloud 2 clusters are lower
than those for the globular clusters.
The main reason for this could be due to the loss of low-mass stars in
globular clusters through dynamical evolution, as pointed out in
\citet{De Marchi2010}.
In other words, the Cloud 2 clusters targeted here are so young that
they has not yet experienced dynamic evolution and they are likely to
keep literally their {\it initial} MF.
Actually, no brown dwarfs have been identified in any globular
clusters to date although some candidates have been found recently
(e.g., \citealt{Dieball2019}, \citealt{Nardiello2023}).
This is because, in addition to their possible loss from globular
clusters due to dynamical evolution, brown dwarfs cannot sustain
hydrogen fusion and they become cooler and fainter with time.
Therefore, our detection of such low-mass sources in clusters with
low metallicity is also one of the significant results of our
observations here.

The ages of the Cloud 2 clusters are estimated to be very young, and
there are very few examples of such young star-forming regions in the
solar neighborhood.
Nevertheless, at least for the currently detected sources, the
color--color diagrams (Figure~\ref{fig:CC_Cloud2cl}) show that the
color excesses from circumstellar materials are very small (see also
\citealt{Yasui2009}),
and there are few sources detected at wavelengths longer than 2
$\mu$m, but not here down to 2 $\mu$m (see \citealt{Izumi2024}).
Therefore, because it is unlikely that the masses of some of the
currently detected or undetected sources will become significantly
larger in the future, the initial mass function will change
drastically in the future for sources produced in the same generation.

For local field stars, there are recent updates with highly accurate IMFs
derived from Gaia data. 
\citet{Hallakoun2021} derived the IMF of stars within 250 pc and in the
mass range of 0.2--1.0 $M_\odot$ using Gaia DR2.
They derived the IMF by classifying stars into four categories:
thin-disc population, thick-disc stars, high-metallicity halo,
and low-metallicity halo ($[{\rm M/H}] < -0.6$). 
For the first three populations, IMFs are found to be similar to
other local regions, with $\log M_C \sim 0.5$ $M_\odot$,
while only low-metallicity halo stars show a bottom-heavy IMF and are
well described by a single power law over most of the mass range,
indicating that $M_C$ is lower than 0.2 $M_\odot$.
The results seem consistent with our results here, in that $M_C$ is
found to be smaller than 0.2 $M_\odot$, although it should be noted
that the low-metallicity halo is likely to be the debris from events,
e.g., accretion by the Milky Way, that occurred very long ago
($\sim$10 Gyr ago). However,
\citet{Li2023} derived the IMF based on $\sim$93,000 spectroscopically
observed M dwarfs (the mass range of 0.3--0.7 $M_\odot$) located
within 100--300 pc, in combination with Gaia DR3, and discussed the
variation of the IMF with metallicity and time.
They suggested that the stellar population that formed at the earliest times
contains fewer low-mass stars compared to the canonical IMF,
independent of stellar metallicity, while 
the fraction of low-mass stars increases with stellar metallicity in
present days.
Their suggestion seems to be consistent with the discussion above
that compares the $M_C$ between old globular clusters and young
open clusters in low-metallicity environments, 
but seems to contradict the comparison of $M_C$
between local star-forming clusters and those in a low-metallicity
environment.
However, since the lower end of the mass ranges in the Gaia studies do
not reach as low as in our study, and the IMF estimated from the Gaia
data does not fully cover the IMF peak, we cannot necessarily compare them on
the same basis.

Theoretically, the temperature of star-forming clouds drops sharply to
a minimum due to the thermal coupling of gas and dust, when
fragmentation tends to occur \citep{{Larson2005}, {Inutsuka1997}}, and
the Jeans mass in these conditions is expected to correspond to the
peak value of the IMF.
At lower metallicities, higher temperatures and hence higher densities
are expected.
For the Jeans mass, $M_J \propto \rho^{-1/2} T^{3/2}$, the effect of
higher density is greater than that of higher temperature, resulting
in lower $M_J$. 
Although this effect has not been explicitly mentioned under the
metallicity of $\sim$$-$1 dex, the plot of fragmentation masses as a
function of heavy elements \citep{Omukai2005} seems to indicate a
trend toward smaller fragmentation masses corresponding to lower
temperature minima with decreasing metallicity.
Recent simulation studies (e.g., \citealt{Bate2019},
\citealt{Chon2021}) also indicate that the peak mass gradually
decreases from $Z/Z_\odot = 10^{-1}$, which is consistent with the
results here,
but the difference in $M_C$ values from those at solar metallicity is
seen in $Z/Z_\odot \lesssim 10^{-4}$, which is much smaller than the
metallicity discussed here.
The derivation of $M_C$ with high accuracy from observations will
help constrain theoretical studies and improve understanding of star
formation in the future.

For star-forming clusters located in the range of $R_G \sim 6$--12
kpc, \citet{Damian2021} derived $M_C$ to study the possible impact of
environmental factors on the form of the IMF at the low-mass end,
using photometry from UKIDSS (United Kingdom Infrared Deep Sky Survey)
and Gaia parallaxes.
They conclude that there is no strong evidence for an environmental
effect in the underlying form of the IMF of their target clusters.
This seems to contradict the results here, but it may be due to the
fact that our results also have large uncertainties, and that their
targets may not have large metallicity differences, or perhaps the
spatial resolution (${\rm FWHM} = 0.8$--1.0 arcsec) and sensitivity
($K \lesssim 18$ mag) of UKIDSS may not be sufficient to resolve
low-mass stars.

The IMFs of young star-forming clusters in low-metallicity environments
have been studied for the outer Galaxy (see Section~\ref{sec:intro}).
Among them, a recent work \citep{Yasui2023}, which is the basis for
the KLF fit in this paper, has for the first time derived detailed
IMFs down to $\simeq$0.1 $M_\odot$ in a low-metallicity environment.
While the mass detection limit was found not to cover the second break
mass of the IMF, the first break masses were estimated to be
$\simeq$0.1 $M_\odot$ for two clusters, which is the upper limit for
$M_C$.
This result is consistent with the $M_C$ derived here for the Cloud 2
clusters, and may support the suggestion here that low $M_C$ values
are characteristics of low-metallicity environments at least in the
Galaxy.
The Large and Small Magellanic Clouds (LMC/SMC) would be a more
typical target for IMF studies in low-metallicity environments
(e.g., \citealt{Sirianni2000}).
However, because mass detection limits had been unable to reach below
$\sim$1 $M_\odot$ until very recently, and even JWST can only reach
down to $\sim$0.1 $M_\odot$ \citep{Leschinski2020} due to the
relatively large distance ($D \simeq 50$ kpc), it is difficult to make
direct comparisons of $M_C$ at this time.


\section{Conclusion} \label{sec:Conclusion}

We have presented NIR observations (from 1 to 2 $\mu$m) using
JWST/NIRCam for Digel Cloud 2, which is a star-forming region in the
Outer Galaxy.
Previous studies found two young open clusters located in the northern
and southern dense cores of the molecular cloud, the Cloud 2-N and -S
clusters, respectively.
The metallicity of Cloud 2 was estimated as $-$0.7 dex.
We derived the substellar mass function for the clusters in this
low-metallicity environment primarily using NIR NIRCam images in the
short wavelength channel (F115W, F150W, and F200W).
The main results can be summarized as follows:

\begin{enumerate}
\item The distance of the Cloud 2 clusters is estimated to be 7.9 kpc
  based on recent results from Gaia DR3.
  Using this distance, the ages of the clusters are estimated as
  $\simeq$0.1 Myr from the projected angular difference of the cluster
  from the current SNR shell front.

\item The FWHM of 0.07 arcsec is achieved for the F200W band images,
  leading to a spatial resolution of 550 au.
  The 10$\sigma$ detection limits, based on pixel-to-pixel noise,
  reach up to $m_{\rm F115W} \simeq 24.5$ mag, $m_{\rm F150W} \simeq
  24.2$ mag, and $m_{\rm F200W} \simeq 23.8$ mag.

 \item The spatial extent of the two clusters are identified from the
   spatial distribution of stellar overdensities detected in the MIRI
   images, which are very sensitive to star formation activity.
   The regions have radii of $\simeq$15 arcsec. 
   Using the ($m_{\rm F150W} - m_{\rm F200W}$) vs. ($m_{\rm F115W} -
   m_{\rm F150W}$) color--color diagram for the detected sources, we
   derived the distributions of reddening, extinction ($A_V$), and
   F200W-band disk excess (F200W excess).
   Based on the resulting distribution of $A_V$, mass-$A_V$-limited
   samples whose low-mass limit is estimated as $\simeq$0.02 $M_\odot$
   for both Cloud 2-N and -S clusters are extracted using the
   color--magnitude diagram.
   Using the mass-$A_V$-limited samples, cluster F200W-band luminosity
   functions (F200W-LFs) are derived with the mass detection limits of
   $\simeq$0.02 $M_\odot$ for both clusters.

   \item Based on the fitting of F200W-LFs by assuming log-normal and
     TPL IMFs, the best derived IMFs are found to cover the the
     characteristic mass of $\log M_C / M_\odot \simeq -1.5$ (0.03
     $M_\odot$).
     This is the first derivation of $M_C$ for young open clusters in
     a low-metallicity environment and was achieved due to the
     high sensitivity of JWST.
     Furthermore, the detection of low-mass sources in the mass range
     of brown dwarfs ($\lesssim$0.08 $M_\odot$) in a low-metallicity
     cluster is an important aspect of this observation because brown
     dwarfs cannot sustain hydrogen fusion and are not found in old
     clusters.

 \item Very similar values of $M_C$ are obtained for the Cloud 2-N
   and -S clusters, suggesting that these values are typical for this
   region.
   It should be noted, however, that the $M_C$ values obtained here
   are subject to various uncertainties in addition to the uncertainty
   from the F200W-LF fitting.

 \item The derived $M_C$ values for the Cloud 2 clusters are generally
   lower than previous derivations for nearby regions.
   In particular, the lower $M_C$ compared to local star-forming open
   clusters, which have similar ages ($\sim$$10^6$--$10^7$ yr) and
   cluster scales ($N_* \lesssim 10^3$), suggests a metallicity
   dependence of $M_C$.
   The lower $M_C$ compared to local globular clusters, which are old
   systems ($\sim$$10^{10}$ yr), suggests that the Cloud 2 clusters are
   very young and have not yet undergone dynamical evolution, where the
   present-day mass function literally reflects the initial mass
   function. 
   Although there appears to be no explicit theoretical mention of $M_C$
   below metallicities of $\sim$$-$1 dex, deriving $M_C$ values with
   high accuracy from observations in such environments would place
   significant constraints on future theoretical studies and improve
   our understanding of star formation.

\end{enumerate}


\begin{acknowledgments}
We thank the anonymous referee for useful comments and suggestions
that helped to improve the manuscript.
We also thank Kazuyuki Omukai and Sunmyon Chon for useful discussions
on theoretical aspects of the IMF in low-metal environments, and Yuji
Ikeda for helpful discussion on statistics.
C.Y. is supported by KAKENHI (18H05441) Grant-in-Aid for Scientific
Research on Innovative Areas. The model fitting was performed using
the Multi-wavelength Data Analysis System operated by the
Astronomical Data Center (ADC) of the National Astronomical
Observatory of Japan.
The work of MER was carried out at the Jet Propulsion Laboratory,
California Institute of Technology, under a contract with the National
Aeronautics and Space Administration (80NM0018D0004).
The data presented in this article were obtained from the Mikulski
Archive for Space Telescopes (MAST) at the Space Telescope Science
Institute. The specific observations analyzed can be accessed via
\dataset[DOI]{https://doi.org/10.17909/4gtt-vx70}. 

\end{acknowledgments}

%

\vspace{5mm}
\facilities{JWST (NIRCam and MIRI), Subaru}


\software{IRAF \citep{{Tody1986}, {Tody1993}}, synphot
  \citep{synphot}, JWST Calibration Pipeline Version
  \citep{Bushouse2023}}



\begin{table*}[!h]
  \caption{The 10$\sigma$ detection limits for JWST/NIRCam
    Observations}
\label{tab:limit}
\begin{center}
\begin{tabular}{lllllllll}
\hline
\hline
Region &
    F115W & F150W & F200W \\
 & (mag) & (mag) & (mag) \\
\hline
Cloud 2-N & 
    24.6 & 24.2 & 23.6 \\
Cloud 2-S &
    24.8 & 24.6 & 23.8 \\
Control & 
    24.6 & 24.6 & 23.8 \\
\hline
\end{tabular}
\end{center}
\end{table*}

\begin{table*}[!h]
  \caption{The best-fit parameters of the Log-normal IMF ($dN / d \log
    m \propto \exp (- C \times (\log m - \log M_C)$) for the Cloud 2-N
    and -S clusters}
\label{tab:KLFfit_LN}
\begin{center}
\begin{tabular}{llll}
\hline
\hline
 Cluster &
    $C$ & $\log M_C$ \\
\hline
Cloud 2-N &
 1.1 [0.1, 4.2] &  $-$1.5 [$-$1.7, $-$1.4] \\
Cloud 2-S & 3.1 [0.9, 6.9] & $-$1.6 [$-$2.0, $-$1.4] \\
 \hline
\end{tabular}

Note. The range of the 68\% confidence level are shown in parentheses.
\end{center}
\end{table*}

\begin{table*}[!h]
  \caption{The best-fit parameters of the TPL IMF ($dN / d \log m
    \propto m^{-\Gamma} \{1 - \exp [- (m / M_C)^\beta]\}$) for the
    Cloud 2-S cluster}
\label{tab:KLFfit_TPL}
\begin{center}
\begin{tabular}{llll}
\hline
\hline
 Cluster &
    $\Gamma$ & $\log M_C$ & $\beta$ \\
\hline
 Cloud 2-S &
0.9 [0.7, 2.0] & $-$1.4 [$-$1.7, $-$1.0] & 1.9 [0.9, 4.0] \\ 
\hline
\end{tabular}

Note. The range of the 68\% confidence level are shown in parentheses.
\end{center}
\end{table*}


\clearpage
\appendix
\section{Color transformations between the NIRCam photometric system and the MKO system}
\label{sec:color_conversion}

Synthetic photometry was generated for the JWST/NIRCam photometric
system, based on \citet{Hewett2006}, who generated synthetic
photometry for the MKO and other filter systems.
The spectra for 151 sources from the Bruzual-Persson-Gunn-Stryker
Atlas \citep{Gunn1983} are used as in \citet{Hewett2006}.
The atlas includes dwarfs (spectral type B9V--M6V), giants
(A3III--M7III), and stars of other luminosity classes (including those
of unknown spectral types).
All magnitude and color calculations are performed using the Python
synphot package \citep{synphot}.
JWST+NIRCam throughput curves (version 5.0: November 2022) were used.
The magnitudes are based on the Vega system by using a reference
spectrum for Vega from \citet{Bohlin2020}.
In addition to the NIRCam filter system, synthetic photometry for the
MKO filter system was generated, according to \citet{Hewett2006}.

\begin{figure}
\begin{center}
  \gridline{\fig{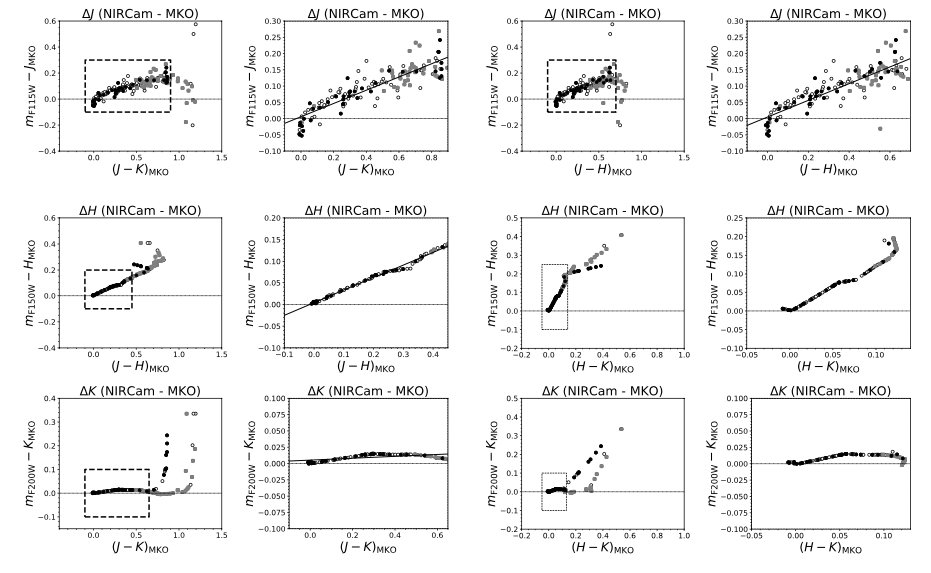}{\textwidth}{}}
  \caption{Differences between the MKO-system and the
    JWST/NIRCam-system magnitudes, as functions of the MKO-system 
    colors.
    In the first and third columns, dwarfs and giants are shown with
    black filled circles and gray filled squares, while the other
    stars are shown with open circles.
    The range where the color-to-magnitude conversion is uniquely
    determined and where the color conversion seems feasible is
    enclosed by a dashed line, and an enlarged view is shown to the
    right of each panel (the second and fourth columns).
    Among these, those with relatively large color ranges (more than
    0.4) are enclosed by thick dashed lines in the first and third
    columns, and the results of fits are shown with thick lines in
    second and fourth columns.} 
\label{fig:CT_JWST_MKO_2MASS}
\end{center}
\end{figure}

Differences between JWST/NIRCam F115W, F150W, and F200W band
magnitudes and MKO-system J, H, and K magnitudes, as a function of MKO
colors are shown in the leftmost panels and third panels from the left
in Figure~\ref{fig:CT_JWST_MKO_2MASS}.
Dwarfs and giants are shown with black filled circles and gray filled
squares, while the other stars are shown with open circles.
As pointed out for the color transformations between the MKO and 2MASS
systems in \citet{Leggett2006},
the same intrinsic dependencies on luminosity class (between red
  giants and dwarfs) in red stars are also seen in the figures.
The range where the color-to-magnitude conversion is uniquely
determined and where the color conversion seems feasible is enclosed
by a dashed line, and an enlarged view is shown to the right of each
panel.
Among these, those with relatively large color ranges (more than 0.4)
are indicated by thick dashed lines, and the color transformations are
derived by linear fitting as follows:

\begin{eqnarray} 
  m_{\rm F115W} - J_{\rm MKO} = (0.204 \pm 0.010) \times (J-K)_{\rm MKO} + (0.006 \pm 0.005) \ {\rm for} \ (J-K)_{\rm MKO} \le 0.9 \label{eq1}\\
  m_{\rm F115W} - J_{\rm MKO} = (0.257 \pm 0.014) \times (J-H)_{\rm MKO} + (0.005 \pm 0.006) \ {\rm for} \ (J-H)_{\rm MKO} \le 0.7 \label{eq2}\\
  m_{\rm F150W} - H_{\rm MKO} = (0.291 \pm 0.003) \times (J-H)_{\rm MKO} + (0.004 \pm 0.001) \ {\rm for} \ (J-H)_{\rm MKO} \le 0.45 \label{eq3}\\
  m_{\rm F200W} - K_{\rm MKO} = (0.014 \pm 0.002) \times (J-K)_{\rm MKO} + (0.005 \pm 0.001) \ {\rm for} \ (J-K)_{\rm MKO} \le 0.65 \label{eq4}
\end{eqnarray}

In Section~\ref{subsec:reduction_photometry}, the objects in the
observation field of view that were also detected with Subaru/MOIRCS
and have colors in the range where color transformations can be
obtained were selected.  The sources are used as standard stars by
obtaining their NIRCam magnitudes using the color transformations from
the MKO to the NIRCam systems.

\section{Color--color diagram in the NIRCam photometric system}
\label{sec:cc_nircam}

\begin{figure}
  \begin{center}
    \gridline{\fig{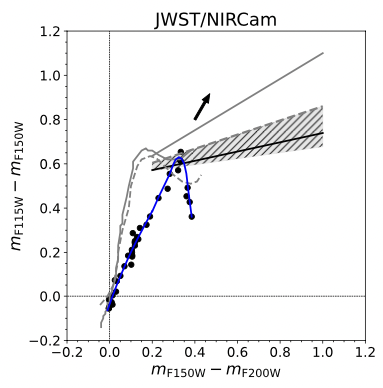}{0.45\textwidth}{}}
    \caption{$(m_{\rm F115W} - m_{\rm F200W})$ vs. $(m_{\rm F115W} -
    m_{\rm F150W})$ color--color diagram in the JWST/NIRCam filter
    system.
   The dwarf stars are shown as black circles, and the dwarf track is
   shown as a blue curve.
   For comparison, the track in the Johnson-Glass system (gray solid
   curve; \citealt{Bessell1988}) and that in the MKO system (gray dashed
   curve; \citealt{Yasui2008}) are shown.
   The area where the CTTS locus is located in the NIRCam system is
   indicated by the hatched area, along with the locus in the CIT
   system (gray solid line; \citealt{Meyer1997}) and that in the MKO
   system (gray dashed line; \citealt{Yasui2008}).
   The arrows indicate the reddening vector for $A_V = 1$ mag from
   \citet{Wang2019}.} 
\label{fig:CC_JWST}
\end{center}
\end{figure}

The CTTS locus was originally derived in the CIT (California Institute
of Technology) system (gray solid line in Figure~\ref{fig:CC_JWST}),
and \citet{Yasui2008} derived the locus in the MKO system (gray dashed
line in Figure~\ref{fig:CC_JWST}) by obtaining CTTS colors from a
2MASS catalog in the 2MASS system and converting them to those in the
MKO system using color transformations by \citet{Leggett2006}.
To derive the CTTS locus in the NIRCam system using the same method,
the color transformations between NIRCam and 2MASS are necessary. 
Using the magnitude differences in the 2MASS and MKO systems presented
by \citet{Hewett2006} and those in the NIRCam and MKO systems obtained
in Appendix~\ref{sec:color_conversion}, we compared the magnitudes in
the NIRCam and 2MASS systems and found that it is difficult to fit for
$m_{\rm F200W} - K$ magnitude differences on the red side ($(J-K)_{\rm
  2MASS} \gtrsim 0.7$ mag or $(H-K)_{\rm 2MASS} \gtrsim 0.1$ mag), as
in the case of magnitude differences in the NIRCam and MKO systems.
Because CTTSs have actual red colors (intrinsic colors $(J-H)_0
\simeq 0.5$--1.0 mag or $(H-K)_0 \simeq 0.2$--1.0 mag), it is
difficult to obtain colors in the NIRCam system by converting from
the color in the 2MASS system.

Instead, we examined the color differences in the NIRCam and MKO
systems to determine the differences in the CTTS locus between the
systems, using the synthetic photometry obtained in
Appendix~\ref{sec:color_conversion}.
Figure~\ref{fig:JH_HK_jwst_mko} shows magnitude comparison
between NIRCam $m_{\rm F115W} - m_{\rm F150W}$ and MKO $J-H$ in the left
panel, and NIRCam $m_{\rm F150W} - m_{\rm F200W}$ and MKO $H-K$ in the
right panel.
Because YSOs without disks are known to show dwarf-like colors (see
Figure~1 in \citealt{Meyer1997} for WTTSs), only dwarfs are shown.
For the comparison of NIRCam $(m_{\rm F115W} - m_{\rm F150W})$ vs. MKO
$J-H$ on all range plots yields $(m_{\rm F115W} - m_{\rm F150W}) =
(0.969 \pm 0.026) \times (J-H)_{\rm MKO} + (-0.012 \pm 0.009)$, shown
with a solid line, indicating that the two colors are the same within
the uncertainties.
For the comparison of NIRCam $(m_{\rm F150W} - m_{\rm F200W})$ vs. MKO
$H-K$ on all range plots yields $(m_{\rm F150W} - m_{\rm F200W}) =
(1.592 \pm 0.048) \times (J-H)_{\rm MKO} + (0.048 \pm 0.007)$.
However, the comparison between the two colors shows a break at $H-K \simeq
0.15$. 
Fitting the plots up to the break ($H-K \le 0.15$) yields $(m_{\rm
  F150W} - m_{\rm F200W})= (2.321 \pm 0.035) \times (H-K)_{\rm MKO} +
(-0.008 \pm 0.003)$ (shown as a dot-dash line).
The plots after the break ($H-K \ge 0.15$) are distributed over a wide
range of horizontal directions, which are located on the left side of
the dashed line showing $m_{\rm F150W} - m_{\rm F200W} = (H-K)_{\rm
  MKO}$.
Therefore, $m_{\rm F150W} - m_{\rm F200W}$ have values between 1 and
2.3 times the MKO $(H-K)$ colors.
In summary, the CTTS locus in the NIRCam is shown with a black line in
Figure~\ref{fig:CC_JWST}, but it can be located in the hatched area
in the figure.
The results are used in Sections~\ref{sec:reddening} to estimate the
reddening for each source and to derive reddening properties for the
Cloud 2 clusters.

In addition, in Figure~\ref{fig:CC_JWST}, the dwarfs in the
JWST/NIRCam filter system from the synthetic photometry are shown as
black dots on the $(m_{\rm F115W} - m_{\rm F200W})$ vs. $(m_{\rm
  F115W} - m_{\rm F150W})$ color--color diagram.
The dwarf star track is shown with a blue curve, which is derived from
eye-fit.
For comparison, the dwarf-star track for the Johnson-Glass system
\citep{Bessell1988} and that for the MKO system \citep{Yasui2008} in the
JHK color--color diagram (H-K and J-H for horizontal and vertical
axes, respectively) are shown with a gray solid line and gray dashed
line, respectively.
In addition, the reddening vector ($A_V = 1$ mag) for the NIRCam
system from \citet{Wang2019} is shown as a black arrow.

\begin{figure}
\begin{center}
  \gridline{\fig{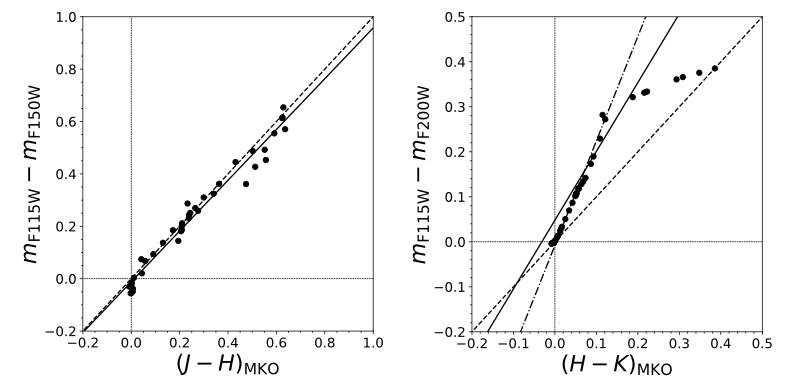}{0.9\textwidth}{}}
  \caption{Comparison of colors in the NIRCam and MKO systems.
    The comparison of NIRCam $(m_{\rm F150W} - m_{\rm F200W})$ and MKO
    $J-H$ are shown in the left panel, while that of $(m_{\rm F150W} -
    m_{\rm F200W})$ vs. MKO $H-K$ is shown in the right panel.
    The dwarfs in the JWST/NIRCam and MKO filter systems from the
    synthetic photometry are shown as black dots.
    The lines show the fits of the plots.
    (See Appendix~\ref{sec:cc_nircam} for details of the fits).} 
\label{fig:JH_HK_jwst_mko}
\end{center}
\end{figure}



\begin{thebibliography}{}
\bibitem[Andersen et al.(2008)]{Andersen2008}
  Andersen, M., Meyer, M.~R., Greissl, J., et al.\ 2008, \apjl, 683, L183.

\bibitem[Bailer-Jones et al.(2021)]{Bailer-Jones2021}
Bailer-Jones, C.~A.~L., Rybizki, J., Fouesneau, M., et al.\ 2021, \aj,
161, 147. 

\bibitem[Baraffe et al.(2015)]{Baraffe2015}
  Baraffe, I., Homeier, D., Allard, F., et al.\ 2015, \aap, 577, A42.

\bibitem[Bate(2019)]{Bate2019}
  Bate, M.~R.\ 2019, \mnras, 484, 2341.

\bibitem[Bessell \& Brett(1988)]{Bessell1988}
Bessell, M.~S., \& Brett, J.~M.\ 1988, \pasp, 100, 1134 

\bibitem[Bohlin et al.(2020)]{Bohlin2020}
  Bohlin, R.~C., Hubeny, I., \& Rauch, T.\ 2020, \aj, 160, 21.

\bibitem[Bouvier et al.(2013)]{Bouvier2013}
  Bouvier, J., Grankin, K., Ellerbroek, L.~E., et al.\ 2013, \aap,
  557, A77. doi:10.1051/0004-6361/201321389

\bibitem[Bushouse et al.(2023)]{Bushouse2023} 
Bushouse, H., Eisenhamer, J., Dencheva, N., et al.\ 2023, Zenodo


\bibitem[Chabrier(2003)]{Chabrier2003}
Chabrier, G.\ 2003, \apjl, 586, L133. doi:10.1086/374879


\bibitem[Chon et al.(2021)]{Chon2021}
  Chon, S., Omukai, K., \& Schneider, R.\ 2021, \mnras, 508, 4175.

\bibitem[Damian et al.(2021)]{Damian2021} Damian, B., Jose, J., Samal, M.~R., et al.\ 2021, \mnras, 504, 2557. 

\bibitem[D'Antona \& Mazzitelli(1997)]{D'Antona1997}
D'Antona, F., \& Mazzitelli, I.\ 1997, Memorie della Societa Astronomica
Italiana, 68, 807

\bibitem[D'Antona \& Mazzitelli(1998)]{D'Antona1998}
D'Antona, F., \& Mazzitelli, I.\ 1998, ASP Conf.~Ser.~134: Brown Dwarfs
and Extrasolar Planets, 134, 442


\bibitem[de Geus et al.(1993)]{de Geus1993}
de Geus, E.~J., Vogel, S.~N., Digel, S.~W., et al.\ 1993, \apjl, 413, L97.

\bibitem[Duch{\^e}ne et al.(2018)]{Duchene2018}
Duch{\^e}ne, G., Lacour, S., Moraux, E., et al.\ 2018, \mnras, 478,
1825. 

\bibitem[De Marchi et al.(2010)]{De Marchi2010}
  De Marchi, G., Paresce, F., \& Portegies Zwart, S.\ 2010, \apj, 718, 105.

\bibitem[Digel et al.(1994)]{Digel1994}
Digel, S., de Geus E. J., \& Thaddeus, P. 1994, ApJ, 422, 92. 

\bibitem[Dieball et al.(2019)]{Dieball2019}
Dieball, A., Bedin, L.~R., Knigge, C., et al.\ 2019, \mnras, 486, 2254.

\bibitem[Elmegreen et al.(2008)]{Elmegreen2008}
Elmegreen, B.~G., Klessen, R.~S., \& Wilson, C.~D.\ 2008, \apj, 681, 365 

\bibitem[Fahrion \& De Marchi(2023)]{Fahrion2023}
  Fahrion, K. \& De Marchi, G.\ 2023, \aap, 671, L14.

\bibitem[Gaia Collaboration et al.(2023)]{Gaia2023}
Gaia Collaboration, Vallenari, A., Brown, A.~G.~A., et al.\ 2023,
\aap, 674, A1.

\bibitem[Gordon et al.(2022)]{Gordon2022}
  Gordon, K.~D., Bohlin, R., Sloan, G.~C., et al.\ 2022, \aj, 163,
  267. 

\bibitem[Gunn \& Stryker(1983)]{Gunn1983}
  Gunn, J.~E. \& Stryker, L.~L.\ 1983, \apjs, 52, 121.

\bibitem[Hallakoun \& Maoz(2021)]{Hallakoun2021}
  Hallakoun, N. \& Maoz, D.\ 2021, \mnras, 507, 398.

\bibitem[Hewett et al.(2006)]{Hewett2006}
Hewett, P.~C., Warren, S.~J., Leggett, S.~K., \& Hodgkin, S.~T.\ 2006,
\mnras, 367, 454  

\bibitem[Hillenbrand(1997)]{Hillenbrand1997}
Hillenbrand, L.~A.\ 1997, \aj, 113, 1733. 

\bibitem[Inutsuka \& Miyama(1997)]{Inutsuka1997}
  Inutsuka, S.-. ichiro . \& Miyama, S.~M.\ 1997, \apj, 480, 681.

\bibitem[Izumi et al.(2017)]{Izumi2017}
Izumi, N., Kobayashi, N., Yasui, C., et al.\ 2017, \aj, 154, 163.

\bibitem[Izumi et al.(2024)]{Izumi2024} 
Izumi, N., Ressler, M.~E., Lau, R.~M., et al.\ 2024, \aj, 168, 68.

\bibitem[Kobayashi \& Tokunaga(2000)]{KT2000}
Kobayashi, N., \& Tokunaga, A.~T.\ 2000, \apj, 532, 423

\bibitem[Kobayashi et al.(2008)]{Kobayashi2008}
Kobayashi, N., Yasui, C., Tokunaga, A.~T., \& Saito, M.\ 2008, \apj, 683, 178

\bibitem[Kraus et al.(2011)]{Kraus2011}
Kraus, A.~L., Ireland, M.~J., Martinache, F., et al.\ 2011, \apj, 731,
8. 

\bibitem[Kroupa(2002)]{Kroupa2002}
Kroupa, P. 2002, Science  295, 82

\bibitem[Lada \& Adams(1992)]{Lada1992} 
Lada, C.~J., \& Adams, F.~C.\ 1992, ApJ, 393, 278 

\bibitem[Lada \& Lada(2003)]{LadaLada2003}
Lada, C.~J., \& Lada, E.~A.\ 2003, \araa, 41, 57 
		
\bibitem[Larson(2005)]{Larson2005}
  Larson, R.~B.\ 2005, \mnras, 359, 211.

\bibitem[Leggett et al.(2006)]{Leggett2006}
Leggett, S.~K., Currie, M.~J., Varricatt, W.~P., et al.\ 2006, \mnras, 373, 781.

\bibitem[Leschinski \& Alves(2020)]{Leschinski2020}
  Leschinski, K. \& Alves, J.\ 2020, \aap, 639, A120.

\bibitem[Li et al.(2023)]{Li2023}
Li, J., Liu, C., Zhang, Z.-Y., et al.\ 2023, \nat, 613, 460.

\bibitem[Lubowich et al.(2004)]{Lubowich2004}
Lubowich, D. A., Brammer, G., Roberts, H.,Millar, T. J., Henkel, C., \&
Pasachoff, J. M. 2004, in Origin and Evolution of the Elements,
ed. A. McWilliam, \& M.  Rauch (Carnegie Observ. 4; Cambridge: Cambridge
Univ. Press), 37

\bibitem[Luhman et al.(2000)]{Luhman2000}
Luhman, K.~L., Rieke, G.~H., Young, E.~T., et al.\ 2000, \apj, 540,
1016.

\bibitem[Meyer et al.(1997)]{Meyer1997}
Meyer, M. R., Calvet, N., \& Hillenbrand, L. A. 1997, AJ, 114, 288

\bibitem[Miller \& Scalo(1979)]{Miller1979}
Miller, G.~E., \& Scalo, J.~M. 1979, ApJS  41, 513

\bibitem[Minowa et al.(2005)]{Minowa2005}
  Minowa, Y., Kobayashi, N., Yoshii, Y., et al.\ 2005, \apj, 629,
  29. 


\bibitem[Moe et al.(2019)]{Moe2019}
  Moe, M., Kratter, K.~M., \& Badenes, C.\ 2019, \apj, 875, 61.
  
\bibitem[Muench et al.(2000)]{Muench2000}
Muench, A. A., Lada, E. A., \& Lada, C. J. 2000, ApJ, 533, 358

\bibitem[Muench et al.(2002)]{Muench2002}
Muench, A.~A., Lada, E.~A., Lada, C.~J., \& Alves, J.\ 2002, \apj, 573, 366

\bibitem[Muench et al.(2003)]{Muench2003}
Muench, A.~A., Lada, E.~A., Lada, C.~J., et al.\ 2003, \aj, 125, 2029 

\bibitem[Muzzio \& Rydgren(1974)]{Muzzio1974}
Muzzio, J. C., \& Rydgren, A. E. 1974, AJ, 79, 864

\bibitem[Nardiello et al.(2022)]{Nardiello2022}
  Nardiello, D., Bedin, L.~R., Burgasser, A., et al.\ 2022, \mnras, 517, 484.

\bibitem[Nardiello et al.(2023)]{Nardiello2023}
  Nardiello, D., Griggio, M., \& Bedin, L.~R.\ 2023, \mnras, 521, L39.

\bibitem[\protect\citeauthoryear{Offner et al.}{2023}]{Offner2023}
Offner S.~S.~R., Moe M., Kratter K.~M., Sadavoy S.~I., Jensen E.~L.~N., Tobin J.~J., 2023, ASPC, 534, 275. 

\bibitem[Omukai et al.(2005)]{Omukai2005}
Omukai, K., Tsuribe, T., Schneider, R., et al.\ 2005, \apj, 626, 627.

\bibitem[Paresce \& De Marchi(2000)]{Paresce2000}
Paresce, F. \& De Marchi, G.\ 2000, \apj, 534, 870. 

\bibitem[Portegies Zwart et al.(2010)]{Portegies Zwart2010}
Portegies Zwart, S.~F., McMillan, S.~L.~W., \& Gieles, M.\ 2010, \araa, 48, 431. 

\bibitem[Rolleston et al.(2000)]{Rolleston2000}
Rolleston, W.~R.~J., Smartt, S.~J., Dufton, P.~L., \& Ryans, R.~S.~I.\
2000, \aap, 363, 537

\bibitem[Ruffle et al.(2007)]{Ruffle2007} 
Ruffle, P.~M.~E., Millar, T.~J., Roberts, H., Lubowich, D.~A., Henkel,
C., Pasachoff, J.~M., \& Brammer, G.\ 2007, \apj, 671, 1766

\bibitem[Rudolph et al.(2006)]{Rudolph2006}
Rudolph, A.~L., Fich, M., Bell, G.~R., et al.\ 2006, \apjs, 162,
346.

  
\bibitem[Russeil et al.(2007)]{Russeil2007}
Russeil, D., Adami, C., \& Georgelin, Y.~M.\ 2007, \aap, 470, 161 

\bibitem[Scalo(1998)]{Scalo1998}
Scalo, J. 1998, in ASP Conf. Ser. 142, The IMF Revisited: A Case for
Variations, ed. G. Gilmore \& D. Howell (San Francisco, CA: ASP), 201

\bibitem[Salpeter(1955)]{Salpeter1955}
  Salpeter, E.~E.\ 1955, \apj, 121, 161.

\bibitem[Siess et al.(2000)]{Siess2000}
Siess, L., Dufour, E., \& Forestini, M.\ 2000, \aap, 358, 593

\bibitem[Simons \& Tokunaga(2002)]{Simons2002}
Simons, D.~A., \& Tokunaga, A.\ 2002, PASP, 114, 169

\bibitem[Sirianni et al.(2000)]{Sirianni2000} 
Sirianni, M., Nota, A., Leitherer, C., De Marchi, G., \& Clampin, M.\
2000, \apj, 533, 203 

\bibitem[Smartt et al.(1996)]{Smartt1996} 
{Smartt, S. J., Dufton, P. L., \& Rolleston, W. R. J. 1996, A\&A, 305, 164}

\bibitem[Smartt \& Rolleston(1997)]{Smartt1997} 
Smartt, S. J., \& Rolleston, W. R. J. 1997, ApJL, 481, L47

\bibitem[Stephens \& Leggett(2004)]{Stephens2004}
Stephens, D.~C., \& Leggett, S.~K.\ 2004, \pasp, 116, 9
		
\bibitem[Stil \& Irwin(2001)]{Stil2001}
Stil, J. M., \& Irwin, J. A. 2001, ApJ, 563, 816. 

\bibitem[Strom et al.(1989)]{Strom1989}
  Strom, K.~M., Strom, S.~E., Edwards, S., et al.\ 1989, \aj, 97,
  1451. 

\bibitem[STScI Development Team (2018)]{synphot}
STScI Development Team 2018, synphot: Synthetic photometry using
Astropy, Astrophysics Source Code Library, ascl:1811.001

\bibitem[Tokunaga et al.(2002)]{Tokunaga2002}
Tokunaga, A. T., Simons, D. A., \& Vacca, W. D. 2002, PASP, 114, 180

\bibitem[Tody(1986)]{Tody1986} Tody, D.\ 1986, \procspie, 627, 733. doi:10.1117/12.968154

\bibitem[Tody(1993)]{Tody1993} Tody, D.\ 1993, Astronomical Data Analysis Software and Systems II, 52, 173

\bibitem[Wall \& Jenkins(2012)]{Wall2012}
  Wall, J.~V. \& Jenkins, C.~R.\ 2012, Practical Statistics for
  Astronomers, by J. V. Wall , C. R. Jenkins, Cambridge, UK: Cambridge
  University Press, 2012

\bibitem[Wang \& Chen(2019)]{Wang2019}
Wang, S. \& Chen, X.\ 2019, \apj, 877, 116. doi:10.3847/1538-4357/ab1c61


  
\bibitem[Yasui et al.(2006)]{Yasui2006} 
Yasui, C., Kobayashi, N., Tokunaga, A.~T., et al.\ 2006, \apj, 649, 753.

\bibitem[Yasui et al.(2008)]{Yasui2008}
Yasui, C., Kobayashi, N., Tokunaga, A.~T., Terada, H., \& Saito, M.\
2008, \apj, 675, 443

\bibitem[Yasui et al.(2009)]{Yasui2009} 
Yasui, C., Kobayashi, N., Tokunaga, A.~T., Saito, M., \& Tokoku, C.\
2009, \apj, 705, 54 

\bibitem[Yasui et al.(2010)]{Yasui2010} 
Yasui, C., Kobayashi, N., Tokunaga, A.~T., Saito, M., \& Tokoku, C.\
2010, \apjl, 723, L113

\bibitem[Yasui et al.(2016a)]{Yasui2016a} 
Yasui, C., Kobayashi, N., Saito, M., et al.\ 2016b, \aj, 151, 115. 

\bibitem[Yasui et al.(2016b)]{Yasui2016b}
Yasui, C., Kobayashi, N., Tokunaga, A.~T., et al.\ 2016a, \aj, 151, 50. 

\bibitem[Yasui et al.(2017)]{Yasui2017}
Yasui, C., Izumi, N., Saito, M., et al.\ 2017, Formation and Evolution
of Galaxy Outskirts, 321, 34. 

\bibitem[Yasui et al.(2021)]{Yasui2021} 
Yasui, C., Kobayashi, N., Saito, M., et al.\ 2021, \aj, 161, 139

\bibitem[Yasui et al.(2023)]{Yasui2023}
Yasui, C., Kobayashi, N., Saito, M., et al.\ 2023, \apj, 943,
137.
		
\end{thebibliography}



\end{document}